\documentclass{aastex631}

\usepackage{amsmath}
\usepackage{CJK}
\usepackage{float}
\usepackage[normalem]{ulem}
\usepackage[shortlabels]{enumitem}
\usepackage{multirow}
\usepackage{graphicx}
\usepackage{xcolor}
\usepackage{upgreek}
\usepackage{amsmath}	
\usepackage{float}

\definecolor{myred}{RGB}{255,66,56} 
\definecolor{myblue}{RGB}{0,0,255}    
\definecolor{mygrey}{RGB}{128,128,128}  
\definecolor{TBR}{RGB}{255,0,255}

\newcommand{\beq}{ \begin{eqnarray} }
\newcommand{\eeq}[1]{\label{#1}\end{eqnarray}}
\newcommand{\eeqn}{ \nonumber \end{eqnarray} }

\newcommand{\Gaia}{{\it Gaia}}

\renewcommand{\tau}{\uptau}

\def\Gaia  {$Gaia$}

\def\h       {\ifmmode{^{\rm h}}\else{$^{\rm h}$}\fi}
\def\m       {\ifmmode{^{\rm m}}\else{$^{\rm m}$}\fi}
\def\s       {\ifmmode{^{\rm s}}\else{$^{\rm s}$}\fi}
\def\deg     {\ifmmode{^{\circ}}\else{$^{\circ}$}\fi}
\def\decdeg  {\ifmmode{{\rlap.}^{\circ}} \else ${\rlap.}^{\circ}$\fi}
\def\decs    {\ifmmode{{\rlap.}^{\rm s}} \else ${\rlap.}^{\rm s}$\fi}
\def\decas   {\ifmmode{{\rlap.}{''}}\else{${\rlap.}{''}$}\fi}
\def\Ho  {\ifmmode{{\rm H}_0}\else{H$_0$}\fi}
\def\Ro  {\ifmmode{{\rm R}_0}\else{R$_0$}\fi}
\def\To  {\ifmmode{\Theta_0}\else{$\Theta_0$}\fi}

\def\Vsbar {\ifmmode {\overline{V_s}}\else {$\overline{V_s}$}\fi}
\def\Usbar {\ifmmode {\overline{U_s}}\else {$\overline{U_s}$}\fi}
\def\Wsbar {\ifmmode {\overline{W_s}}\else {$\overline{W_s}$}\fi}

\def\mux    {\ifmmode {\mu_x}\else {$\mu_x$}\fi}
\def\muy    {\ifmmode {\mu_y}\else {$\mu_y$}\fi}
\def\mura   {\ifmmode {\mu_{\alpha}}\else {$\mu_{\alpha}$}\fi}
\def\mude   {\ifmmode {\mu_{\delta}}\else {$\mu_{\delta}$}\fi}

\def\gax{\mathrel{\rlap{\lower4pt\hbox{\hskip1pt$\sim$}}
    \raise1pt\hbox{$>$}}}

\def\d    {\ifmmode {{\rlap{.}}^\circ}\else {${\rlap{.}}^\circ$}\fi}
\def\s    {\ifmmode {{\rlap{.}}^s}\else {${\rlap{.}}^s$}\fi}
\def\as   {\ifmmode {{\rlap{.}}^{''}}\else {${\rlap{.}}^{''}$}\fi}

\def\lax{\mathrel{\rlap{\lower4pt\hbox{\hskip1pt$\sim$}}
    \raise1pt\hbox{$<$}}}   
\def\gax{\mathrel{\rlap{\lower4pt\hbox{\hskip1pt$\sim$}}
    \raise1pt\hbox{$>$}}}   
    
\def\aap{A\&A}

\def\apjs{ApJS}
\def\apjl{ApJ}

\shorttitle{GeoVLBI at 22/43/88/132 GHz}
\shortauthors{Xu et al.}
\graphicspath{{./}{figures/}}

\begin{document}
\begin{CJK*}{UTF8}{gbsn}
\title{A Geodetic and Astrometric VLBI Experiment at 22/43/88/132 GHz}

\correspondingauthor{Shuangjing Xu}
\email{sjxuvlbi@gmail.com}
\author[0000-0003-2953-6442]{Shuangjing Xu} 
\affiliation{Korea Astronomy and Space Science Institute, 776 Daedeok-daero, Yuseong-gu, Daejeon 34055, Republic of Korea}

\author[0000-0001-7003-8643]{Taehyun Jung} 
\affiliation{Korea Astronomy and Space Science Institute, 776 Daedeok-daero, Yuseong-gu, Daejeon 34055, Republic of Korea}

\author[0000-0003-1353-9040]{Bo Zhang} 
\affiliation{Shanghai Astronomical Observatory, Chinese Academy of Sciences, 80 Nandan Road, Shanghai 200030, People's Republic of China}

\author[0000-0001-9602-9489]{Ming Hui Xu} 
\affiliation{GFZ German Research Centre for Geosciences, 14473 Potsdam, Germany}
\affiliation{Aalto University Metsähovi Radio Observatory, Metsähovintie 114, 02540 Kylmälä, Finland}

\author[0000-0003-1157-4109]{Do-Young Byun} 
\affiliation{Korea Astronomy and Space Science Institute, 776 Daedeok-daero, Yuseong-gu, Daejeon 34055, Republic of Korea}

\author[0009-0005-7123-4378]{Xuan He} 
\affiliation{Shanghai Astronomical Observatory, Chinese Academy of Sciences, 80 Nandan Road, Shanghai 200030, People's Republic of China}
\affiliation{University of Chinese Academy of Sciences, No.19 (A) Yuquan Rd, Shijingshan, Beijing 100049, People's Republic of China}

\author[0000-0002-5814-0554]{Nobuyuki Sakai} 
\affiliation{National Astronomical Research Institute of Thailand (Public Organization), 
260 Moo 4, T. Donkaew, A. Maerim, Chiang Mai, 50180, Thailand}

\author[0000-0003-1751-676X]{Oleg Titov} 
\affiliation{Geoscience Australia, PO Box 378, Canberra 2601, Australia}

\author[0000-0001-7308-6659]{Fengchun Shu} 
\affiliation{Shanghai Astronomical Observatory, Chinese Academy of Sciences, 80 Nandan Road, Shanghai 200030, People's Republic of China}

\author{Hyo-Ryoung Kim} 
\affiliation{Korea Astronomy and Space Science Institute, 776 Daedeok-daero, Yuseong-gu, Daejeon 34055, Republic of Korea}

\author{Jungho Cho} 
\affiliation{Korea Astronomy and Space Science Institute, 776 Daedeok-daero, Yuseong-gu, Daejeon 34055, Republic of Korea}

\author{Sung-Moon Yoo} 
\affiliation{Korea Astronomy and Space Science Institute, 776 Daedeok-daero, Yuseong-gu, Daejeon 34055, Republic of Korea}

\author{Byung-Kyu Choi} 
\affiliation{Korea Astronomy and Space Science Institute, 776 Daedeok-daero, Yuseong-gu, Daejeon 34055, Republic of Korea}

\author[0000-0001-5020-8684]{Woo Kyoung Lee} 
\affiliation{Korea Astronomy and Space Science Institute, 776 Daedeok-daero, Yuseong-gu, Daejeon 34055, Republic of Korea}

\author[0000-0002-8604-5394]{Yan Sun}
\affiliation{Shanghai Astronomical Observatory, Chinese Academy of Sciences, 80 Nandan Road, Shanghai 200030, People's Republic of China}

\author[0000-0001-7573-0145]{Xiaofeng Mai}
\affiliation{Shanghai Astronomical Observatory, Chinese Academy of Sciences, 80 Nandan Road, Shanghai 200030, People's Republic of China}
\affiliation{University of Chinese Academy of Sciences, No.19 (A) Yuquan Rd, Shijingshan, Beijing 100049, People's Republic of China}

\author{Guangli Wang} 
\affiliation{Shanghai Astronomical Observatory, Chinese Academy of Sciences, 80 Nandan Road, Shanghai 200030, People's Republic of China}

\begin{abstract}
Extending geodetic and astrometric Very Long Baseline Interferometry (VLBI) observations from traditional centimeter wavebands to millimeter wavebands offers numerous scientific potentials and benefits. However, it was considered quite challenging due to various factors, including the increased effects of atmospheric opacity and turbulence at millimeter wavelengths.
Here, we present the results of the first geodetic-mode VLBI experiment, simultaneously observing 82 sources at 22/43/88/132 GHz (K/Q/W/D bands) using the Korean VLBI Network (KVN). We introduced the frequency phase transfer (FPT) method to geodetic VLBI analysis, an  approach for calibrating atmospheric phase fluctuations at higher frequencies by transferring phase solutions from lower frequencies. 
With a 2-minute scan, FPT improved the signal-to-noise ratio (SNR) of most fringes, some by over $100\%$ , thereby enhancing the detection rate of weak sources at millimeter wavebands.
Additionally, FPT reduced systematic errors in group delay and delay rate, with the weighted root-mean-squares (WRMS) of the post-fitting residuals decreasing from 25.0 ps to 20.5 ps at the W band and from 39.3 ps to 27.6 ps at the D band. 
There were no notable differences observed in calibrating atmospheric phase fluctuations at the K band (WRMS = 12.4 ps) and Q band (WRMS = 11.8 ps) with the KVN baselines.
This experiment demonstrated that the millimeter waveband can be used for geodetic and astrometric applications with high precision.
\end{abstract}


\section{Introduction} \label{sec:intro}

Geodetic and astrometric Very Long Baseline Interferometry (VLBI) observations have made significant contributions to astronomy and geodesy over the past 40 years, particularly in the areas of the terrestrial reference frame (TRF), the celestial reference frame (CRF), and the earth orientation parameters (EOPs) \citep{1998RvMP...70.1393S}. The basic concept of geodetic and astrometric VLBI involves using pairs of radio telescopes to observe the signals of compact extra-galactic objects that emit radiation in the radio frequency regime. By analyzing the differences in the arrival times of the same wavefront between two telescopes, geodetic parameters such as telescope coordinates, positions of celestial objects, and EOPs can be inferred for various scientific and practical applications \citep{Nothnagel2019}. 
In addition to the traditional S/X band (2.3/8.4 GHz), multiple frequency bands have recently been employed to enhance the potential of geodetic VLBI observations.

The accuracy of geodetic and astrometric measurements relies on the precision of derived group delays, baseline length, and systematic delay errors. 
The broadband geodetic VLBI system at 2-14 GHz, known as VLBI Global Observing System (VGOS), has improved the precision of group delays to a few picoseconds (ps; 1 ps = $10^{-12}$ s) \citep{2018RaSc...53.1269N}, however, uncompensated systematic errors at the level of 20 ps  still dominate the error budget \citep{2021JGeod..95...51X}.
Further efforts are required to calibrate systematic errors originating from tropospheric delay \citep{2024arXiv240408800P} and source structure \citep{2022A&A...663A..83X} to achieve the VGOS's goal of 1 mm position accuracy.

Higher frequency bands offer advantages such as achieving higher-resolution imaging of radio sources, mitigating source structure effects \citep{2023AJ....165..139D}, measuring frequency-dependent position shifts in Active Galactic Nuclei (AGN) jets (i.e., core-shift) \citep{1979ApJ...232...34B,2011Natur.477..185H}, 
and reducing interference from scattering \citep{2022MNRAS.515.1736K} and ionospheric plasma effects  \citep{2009ApJ...699.1395F}.  
The K band (24 GHz)， X/Ka band (8.4/32 GHz), and Q band  (43 GHz) \citep{2010AJ....139.1695L,2020A&A...644A.159C} 
have been used for establishing a multi-frequency International Celestial Reference Frame (ICRF).
In the meantime, the \Gaia\ satellite mission realized the first extra-galactic frame at optical wavelengths \citep{2022A&A...667A.148G}. 
How to realize a fully consistent and integrated multi-waveband celestial reference frame becomes an important issue \citep{Charlot2022}.
In addition, the use of millimeter wavebands for geodesy is valuable for determining the station coordinates of  antennas without receivers operating at lower frequencies.

The independent geodetic VLBI programs operating at different frequencies  may have astrometric limitations in detecting the core-shift in most ICRF sources \citep{2024arXiv240408800P}. 
As frequency increases, the signal-to-noise ratio (SNR) of VLBI fringes is affected by decreased flux densities of sources, shorter coherence times, increased atmospheric absorption, and higher receiver temperatures, limiting the precision of group delays at millimeter wavelengths.
The single K band observation also has difficulty in calibrating the ionospheric delay \citep{2010AJ....139.1695L}. 

The Korean VLBI Network (KVN) has the capability of simultaneously observing at multiple frequencies
\citep{2013PASP..125..539H}, including K band at 18-26 GHz, Q band at 35-50 GHz, W band at 85-116 GHz, and D band\footnote{
The D band refers to an industrial waveguide band designated for 110-170 GHz. 
The Institute of Electrical and Electronics Engineers (IEEE) designates the frequency range 110 to 300 GHz as the Millimeter band.
} at 125-142 GHz.
A similar K/Q/W band system is being developed globally
\citep{2023arXiv230604516D}.
This kind of system is
particularly useful for extending the coherence time at millimeter wavelengths using the frequency phase transfer (FPT) technology \citep{2020A&ARv..28....6R}. 
Specifically, it achieves this by calibrating the tropospheric phase at higher frequencies through transferring phase solutions from lower frequencies. This enables the observation of more sources and facilitates the measurement of core shift effects using source-frequency phase-referencing (SFPR) astrometry \citep{2015AJ....150..202R,2015JKAS...48..277J}.
The global K/Q/W band  system may also benefit the ICRF by employing geodetic and astrometric VLBI across a broad frequency range from 20 to over 100 GHz. This approach offers several advantages:
1) simultaneous multi-frequency investigation of core shift effects and source structures;
2) monitoring numerous ICRF source images with resolutions of a few tens of microarcseconds;
3) overcoming limitations in group delay precision at millimeter wavelengths through very broad bandwidth synthesis.

In this paper, we present the first geodetic VLBI observation at 22/43/88/132 GHz simultaneously using the KVN as a pilot experiment for future broad bandwidth synthesis 
from 20 to over 100 GHz.

\section{Observations and general DATA ANALYSIS}

\begin{deluxetable*}{crlrrrc}
\tablecaption{KVN geodetic observation in 2021 December 07/08\label{tab:obs1}}
\tablewidth{0pt}
\tablehead{
\colhead{Band} & \colhead{Frequency}  & \colhead{Polarization\tablenotemark{a}} & \colhead{Channel} & \colhead{Number  of} & \colhead{Scheduled} & \colhead{Scheduled}  \\
\colhead{ID}   & \colhead{(MHz)}  & \colhead{}     & \colhead{bandwidth (MHz)}        & \colhead{channel}          & \colhead{scans}                          & \colhead{observables}                         
}
\startdata
K & 21984 & RCP  &   512   &  1   & 485  & 1455  \\
Q & 42620 & LCP  &   512   &  1   & 485  & 1455   \\
W & 87936 & LCP  &   512   &  1   & 485  & 1455    \\
D &131904 & LCP  &   512   &  1   & 485  & 1455    \\
\enddata
\tablenotetext{a}{RCP is right circular polarization, and LCP is left  circular polarization.}
\end{deluxetable*}

We conducted the first geodetic and astrometric VLBI experiment observed at K/Q/W/D bands (22/43/88/132 GHz) simultaneously using the KVN under the East Asian VLBI Network (EAVN) \citep{2022Galax..10..113A} program a2129a
.
The KVN consists of three 21-m antennas: KVN-Yonsei (KYS), KVN-Ulsan (KUS), and KVN-Tamna (KTN), with baseline lengths ranging from 305 to 476 km.
We used  a 24-hour track for the session from 2021-Dec-07/15:25:00 to 2021-Dec-08/15:25:00.
The received signals were recorded with four 512 MHz base-band channels (BBCs) and recorded right (for K band) or left (for Q/W/D bands) circularly polarized signals with Nyquist sampling and 2 bits per sample for a total sampling rate of 8 Gbps.
A summary of the observation is listed in Table \ref{tab:obs1}.
 
\subsection{Scheduling}

There are a few known sources at W and D bands that can be used for geodetic VLBI observations. Fortunately, the Multi-frequency AGN Survey with the KVN (MASK) project
\citep{2018evn..confE.104J}
provided suitable candidates for this geodetic session.
MASK is a simultaneous multi-frequency VLBI fringe detection survey utilizing the FPT technique for $\sim$1533 AGN samples from the KVN single dish survey \citep{2017ApJS..228...22L} with a bandwidth of 64 MHz for each K/Q/W/D band (1 Gbps in total) and a 30-min scan for each source. 
MASK has detected hundreds of AGNs at W and D bands using the  phase self-calibration or the FPT technique (Jung et al. in preparation). We selected a candidate catalog with 125 sources, all of which were detected at K/Q/W/D bands with MASK and included in the ICRF3 K band catalog \citep{2020A&A...644A.159C}.

Observation scheduling was performed using the NASA VLBI scheduling program {\it SKED} \citep{gipson2018sked}. 
We used a 2-min scan for each source with 8 Gbps (2 Gbps per band) recording mode (with low system noise temperature in winter), which has comparable sensitivity to the MASK project (30-min scan and 1 Gbps recording mode). 
The best 80 targets were selected automatically by {\it SKED}.
During the geodetic observations, we also included five $\sim$8-min ``phase-referencing blocks'' for two source pairs: 4C39.25$\sim$0945+408 (4.5\deg\ separation) and 2136+141$\sim$2150+173 (4.5\deg\ separation). For each ``phase-referencing block'', we switched between the source pairs for 3 cycles with a switching time of $\sim$20 s and on-source time of 60 s for each target. (The results of the phase-referencing blocks are not within the scope of this paper and will be reported elsewhere.)
The sources 4C39.25 and 2136+141 were included in the best 80 targets, and the sources 0945+408 and 2150+173 were added manually. All 82 sources are listed in  APPENDIX \ref{app}.

A total of 485 scans were scheduled for each KVN baseline using {\it SKED}.
Since the KVN has its own field system supporting the NRAO VLBI scheduling program {\it SCHED}, the observing scans from {\it SKED} were modified to {\it SCHED} format.

\subsection{Correlation and Fringe fitting}\label{sec:corr}

The observations were correlated using the {\it DiFX} software correlator \citep{2011PASP..123..275D} in Daejeon, South Korea. The output of the correlator was converted to {\it Mark 4} format in order to be compatible with the Haystack Observatory Processing System ({\it HOPS}) suite of programs \citep{2022ascl.soft05019W} and converted to {\it FITS} format for imaging with NRAO Astronomical Image Processing System
({\it AIPS}) \citep{2003ASSL..285..109G}. 

The {\it HOPS} main tool, {\it fourfit}, was used to estimate group and phase delays, phase, delay rate, and cross-correlation amplitude for each observation using the {\it Mark 4} data. 
As the hardware phase calibration system to cover the entire frequency range of KVN (18 - 142 GHz) is under development, we implemented manual phase calibration through a single scan of the bright source OJ287, to align the delays and phases among different frequency bands. Contrary to the standard practices in geodetic VLBI, we introduced the utilization of FPT to address the effects of atmospheric  turbulence in millimeter wavebands. This method is elucidated in Section \ref{sec:FPT}.
A database in {\it vgosDb} format \citep{2016ivs..conf..222B}
was finally  produced in three steps: ``{\it vgosDbMake}'' produced the skeleton database for all observables (version 1); ``{\it vgosDbCalc}'' added the apriori values and partial derivatives (version 2), and ``{\it vgosDbProcLogs}'' added meteorological information (version 3; no cable cal in KVN log file)\citep{2014ivs..conf..253B}.

\subsection{Geodetic data analysis} \label{sec:analysis}

The geodetic analysis was conducted using the {\it nuSolve} program \citep{2014ivs..conf..253B}. This program operates on the {\it vgosDb} database to perform least-squares estimation of various geodetic, geophysical, astronomical, and instrumental parameters.  
We disassembled the database into individual bands (e.g., K, Q, W, D) and dual-bands (e.g., Q/K, W/K, D/K)  by manually editing the ``wrapper'' file in the {\it vgosDb} database.

Obtaining the final geodetic estimate utilized multiple steps proceeding  from  least precise to most precise. The detailed steps can be found in the  {\it User Guide of nuSolve}
and \cite{2021JGeod..95...65N}. 
External files containing a priori information are used in the analysis, such as updated station coordinates from the multi-epoch EAVN geodetic observations \citep{2021evga.conf...71X}, source positions from the ICRF3 K band catalog \citep{2020A&A...644A.159C}, and earth rotation parameters from the VLBI solution provided by NASA Goddard Space Flight Center (GSFC).
The group delays with the single 512 MHz channel in our data are unambiguous.
We start with simple parameterization, only clock shifts and rates, and perform an analysis of the group delays. At this stage, we use the KYS station as the clock reference and have not identified any clock breaks at any of the stations.
Then we add zenith delays and station positions to the list of estimated parameters. 

Also, we test ionospheric corrections for dual-frequency (Q/K, W/K, D/K) band data.
Usually, geodetic VLBI uses multi-frequency channels for each frequency band, i.e., the bandwidth synthesis technique, to improve the group-delay measurement precision. In this case, an ``effective frequency'' is calculated and assigned to ionosphere group delays as the approximate reference frequency \citep{2013aesg.book.....B}. 
We calculated the central frequencies of each band (22240 MHz for K band, 42876 MHz for Q band, 88192 MHz for W band, 132160 MHz for D band) as the effective frequency and used them latter in ionospheric calibration.

Time-varying models of clock and tropospheric parameters are introduced in the latter stage. They are modeled as continuous piece-wise linear (PWL) functions with incremental rates. For such a PWL model, the estimated values are, for each parameter, an initial value and rate for the first interval and a new rate for each of the successive equal-duration intervals. The results reported in the remainder of this paper are based on a PWL interval of 30 min for the troposphere and 60 min for the clock. 
The daily averaged atmospheric gradients \citep{1995GeoRL..22.1041M} are estimated
with constraints in this independent solution.
In the last stage of data processing, additional parameters such as the rate of Earth rotation and angles of nutation  are included. We also re-weight the observations by examining the additive noise required to achieve a chi-squared per degree of freedom (chi2pdof) of approximately 1. Any post-fit delay residuals greater than 3.5 times their re-weighted uncertainty are marked for exclusion. We iterate through the estimation, re-weighting, and outlier-check sequence until no outliers are detected, resulting in the final solution.
To efficiently compare results from different frequency bands and/or parameters, we use the script mode with {\it nuSolve} \citep{2023ivs..conf..159B}. 

\section{Application of FPT in geodetic VLBI }\label{sec:FPT}

The inherent challenges of geodetic VLBI in millimeter wavebands, such as sensitivity, can be effectively addressed due to the impressive performance of the KVN telescopes \citep{2014AJ....147...77L},
characterized by their high aperture efficiency, precise pointing accuracy, receivers with low noise temperature, wide-band digital backend, and rapid slewing speed.  
Of particular significance is the telescopes' ability to operate at multiple frequencies, which enables us to mitigate atmospheric phase fluctuations using the  FPT method.
In the FPT method, high-frequency (target frequency) observations are calibrated using scaled solutions obtained from a lower, more easily manageable frequency \citep{2011PASJ...63..375J,2020A&ARv..28....6R}.

The majority of mm-VLBI observations serve imaging purposes, wherein the 
rapid nonlinear phase (atmospheric phase fluctuations) is usually estimated per scan using self (on-source) detections from a single reference station to other stations \citep{2019ApJ...882...23B}.  The correlated signal must have a high SNR so that the atmospheric phase can be estimated on a short timescale (a few seconds).  
Therefore, this method is limited to using bright sources or using a high-sensitivity station as the phase reference.  In addition, it is difficult to distinguish 
nonlinear atmospheric phase from the linear phase drift due to delay rate, and
can lead to the loss of frequency-dependent information in the phases.
The FPT method can be effectively employed to overcome these limitations, either simultaneously or through fast frequency switching.

\begin{figure}[htb!]
\epsscale{1}
\plotone{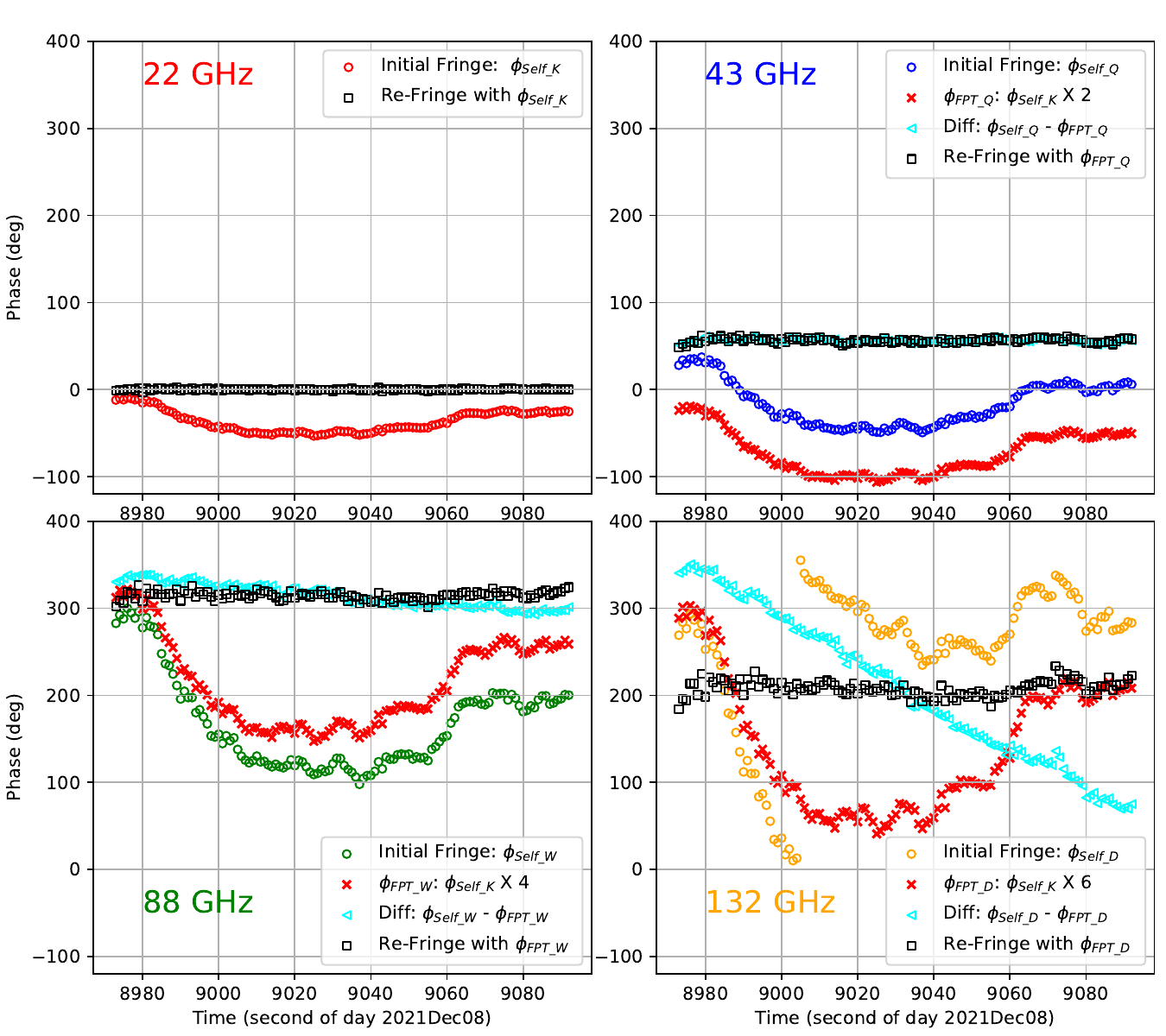}
\caption{
Comparison atmospheric phase fluctuations with self detection ($\phi_{Self}$) and FPT ($\phi_{FPT}$) method using a strong source 3C279 on KTN-KYS baseline. 
The four panels represent distinct frequency bands. 
The 2$\pi$ ambiguity of phase is ignored for clarity.}
\label{fig:ftp}
\end{figure}

\begin{figure}[htb!]
\epsscale{1}
\plotone{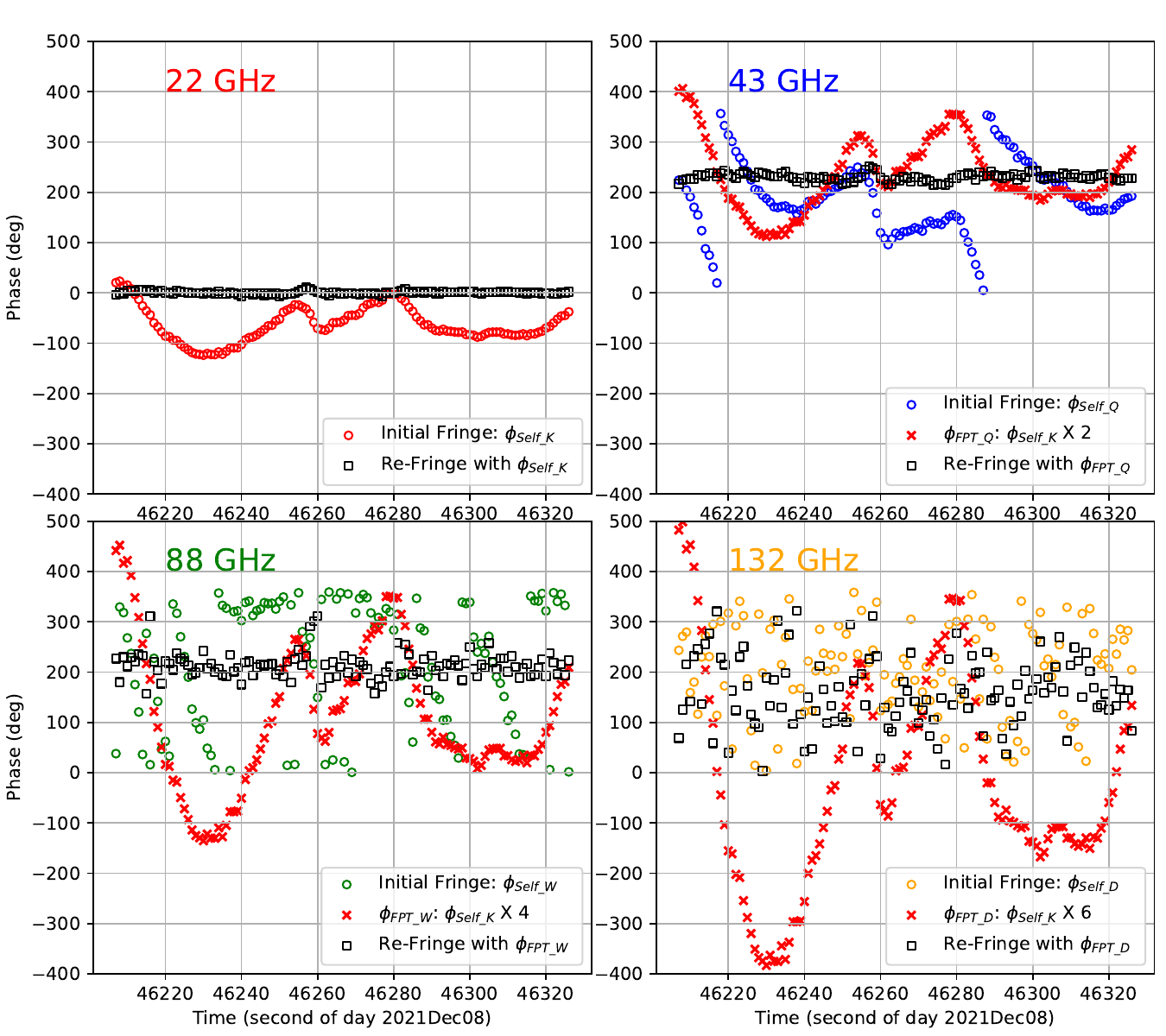}
\caption{
Same as Figure \ref{fig:ftp}, but for a weaker source 0642$+$449. 
}
\label{fig:ftp437}
\end{figure}

We effectively employed the FPT method using the packets  {\it HOPS} for geodetic data,  drawing from the guidance provided in the {\it fourfit} user's manual and a tutorial \citep{Fish2015}.
This implementation involved the following steps:
\begin{itemize}
  \item   Initiating fringe fitting using the {\it fourfit} program and employing the {\it alist} program to track the fringe phase,  SNR, and other information for each observation, leading to the creation of a corresponding text file. 
  \item Employing the {\it fringex} program to segment data from a single scan time  ($\sim$2 min) into a few seconds. This step utilized the aforementioned text file ({\it alist}) as input and provided atmospheric phase fluctuations using self detection at each bands (e.g., $\phi_{Self\_K}$, $\phi_{Self\_Q}$, $\phi_{Self\_W}$, and $\phi_{Self\_D}$).
Choosing  a reference station and  an appropriate cadence for segmented phases is crucial; it should be short enough to capture the atmospheric phase fluctuations while maintaining enough SNR. 
In this experiment, we used KYS as the reference station and examined cadences of 1 second, 3 seconds, 5 seconds, and 10 seconds. Subsequently, we selected the smallest cadence with an SNR exceeding 10. Given the high SNRs in the K band,  $\sim$90$\%$ of the scans adopted a 1-second cadence. However, for a small number  of weaker scans with an SNR below 10 at the 10-second cadence, we refrained from applying FPT to prevent the potential deterioration of results.
  \item   Calculating atmospheric phase fluctuations at higher frequencies by transferring phase solutions from a lower frequency (i.e., K band), based on frequency ratios. As illustrated in Table \ref{tab:obs1}, the frequency ratios are 1.93 (approximately equal to 2) for Q/K bands, 4 for W/K bands, and 6 for D/K bands, resulting in $\phi_{FPT_Q} = \phi_{Self_K} \times 2$, $\phi_{FPT_W} = \phi_{Self_K} \times 4$, and $\phi_{FPT_D} = \phi_{Self_K} \times 6$. 
  We have applied phases from one polarization (LCP of K band) to the other (RCP of Q/W/D band), as atmospheric phases remain unaffected by polarization.
  \item Within the {\it HOPS} framework, the solutions were based on baselines. We designated KYS as the reference station and ``viewed'' the atmospheric phases ($\phi_{Self\_K}$, $\phi_{FPT\_Q}$, $\phi_{FPT\_W}$, $\phi_{FPT\_D}$) originating from KUS and KTN stations. Finally, we integrated this ``{\it ad hoc}" phase information into {\it fourfit} and re-performed the process of fringe fitting.
  \end{itemize}
  
As shown in Figure \ref{fig:ftp} and \ref{fig:ftp437}, the FPTed phases  exhibit close agreement with the trends observed in self detected phases at each K/Q/W/D bands. The clear linear phase differences ($Diff:\phi_{Self}-\phi_{FPT}$ in Figure \ref{fig:ftp}) between self detection and FPT method indicate the different delay rates.
In Table \ref{tab:fpt}, it is noteworthy that the fringe quality (``Qcodes'') of FPT-derived detection is comparable to those of self-detection.
The term ``Qcodes'' represents the fringe quality code as defined by {\it HOPS-fourfit}, where higher Qcodes indicate better quality. 
Subsequent to the application of FPT, the proportion of all usable observables with Qcodes of 9 has improved from  79$\%$  to 99$\%$, characterized by heightened SNR and improved delay/rate precision.
The improvements in group delay and rate accuracy are presented in Section \ref{sec:results}. 
After applying FPT, the presence of phase differences among different bands in Table \ref{tab:fpt} necessitates further investigation through precise instrumental phase calibration. However, their influence on group delay and delay rate measurements is negligible in the following analysis.

\begin{deluxetable*}{rrrlrrrrrcc}[htb!]
\tablecaption{The fringe information with observables in Figure \ref{fig:ftp} and \ref{fig:ftp437} 
\label{tab:fpt}}
\tablewidth{0pt}
\tablehead{
\colhead{Source} & \colhead{Band} & \colhead{Frequency}  & \colhead{Atmospheric phase}  & \colhead{Qcodes\tablenotemark{a}}   & \colhead{SNR}   & \colhead{Phase} & \colhead{Group delay}  & \colhead{Delay rate}  &  \colhead{}  & \colhead{}  \\
\colhead{} & \colhead{ID} & \colhead{(MHz)}  & \colhead{calibration}  & \colhead{}  & \colhead{} & \colhead{(degree) }& \colhead{precision (ps)} & \colhead{precision (ps/s)}& \colhead{}
}
\startdata
3C279   & K & 21984  & NO    & 8 &  5108.0 &  -36.7 &  2.1E-01   & 4.0E-5 \\
  &    &        & $\phi_{Self\_K}$ & 9 &  5233.1 &    0.1 &  2.1E-01   & 3.9E-5 \\
  &  Q & 42620  & NO    & 7 &  3853.5 &  -17.3 &  2.8E-01   & 2.8E-5 \\
  &    &        & $\phi_{Self\_Q}$  & 9 &  4256.0 &    -0.1 &  2.5E-01   & 2.5E-5 \\
  &    &        & $\phi_{FPT\_Q}$   & 9 &  4253.4 &   56.5 &  2.5E-01   & 2.5E-5 \\
  &  W & 87936  & NO    & 5 &   740.7 &  158.0 &   1.5E+00  & 7.1E-5 \\ 
  &    &        & $\phi_{Self\_W}$  & 9 &  1121.8 &    0.3 &   9.6E-01  & 4.7E-5 \\
  &    &        & $\phi_{FPT\_W}$   & 9 &  1120.3 &  -45.5 &   9.6E-01  & 4.7E-5 \\
  &  D &131904  & NO    & 5 &   275.1 &  -75.7  &  3.9E+00  & 1.3E-4 \\
  &    &        & $\phi_{Self\_D}$  & 9 &   394.5 &  1.1  &  2.7E+00  & 8.8E-5 \\
  &    &        & $\phi_{FPT\_D}$   & 9 &   392.3 &  -152.8 &  2.7E+00  & 8.9E-5 \\
\hline
0642+449 & K & 21984  & NO    & 6 &  349.7 &  -64.1 &  3.1E+00   & 5.9E-4 \\
  &    &        & $\phi_{Self\_K}$ & 9 &  412.2 &  0.3 &  2.6E+00   & 4.9E-4 \\
  &  Q & 42620  & NO    & 5 &  131.9 &  -175.0 &  8.2E+00   & 8.2E-4 \\
  &    &        & $\phi_{FPT\_Q}$   & 9 &  254.2 &  -129.9 &  4.2E+00   & 4.3E-4 \\
  &  W & 87936  & NO    & 9 &   18.7 &  -17.8 &   5.7E+01  & 2.8E-3 \\ 
  &    &        & $\phi_{FPT\_W}$   & 9 &  37.2 &  -150.0 &   2.9E+01  & 1.4E-3 \\
  &  D &131904  & NO    & 0 &   5.3 &  -127.5  &  2.0E+02  & 6.5E-3 \\
  &    &        & $\phi_{FPT\_D}$   & 9 &   9.2 &  156.7 &  1.2E+02  & 3.8E-3 \\
\enddata
\tablenotetext{a}{The ``Qcodes'' represents the fringe quality code as defined by {\it HOPS-fourfit}, where higher Qcodes indicate better quality. }
\end{deluxetable*}

\section{Results}\label{sec:results}

The weather conditions were clear during the observation period, with median  system temperature (Tsys) ranges as follows: 87-131 K for the K band, 96-103 K for the Q band, 189-193 K for the W band, and 186-270 K for the D band  at the three KVN stations.
These favorable conditions have led to excellent detection (SNR $>=$ 7), with the detection rate of 99.8$\%$ at K band,  99.8$\%$ at Q band,  95.5$\%$ at W band (or 96.3$\%$ with  FPT), and 68.2$\%$ (or 70.9$\%$  with FPT) at D band for 485 scans on KYS-KTN baseline. 
However, the Q band data from the KUS station exhibited a rare lower SNR compared to other stations, for unknown instrumental reasons at this time. Therefore, we didn't use the Q band data of KUS station in the final results.

\subsection{The group delay and delay rate measurements at 22/43/88/132 GHz}

\begin{figure}[htb!]
\epsscale{1}
\plotone{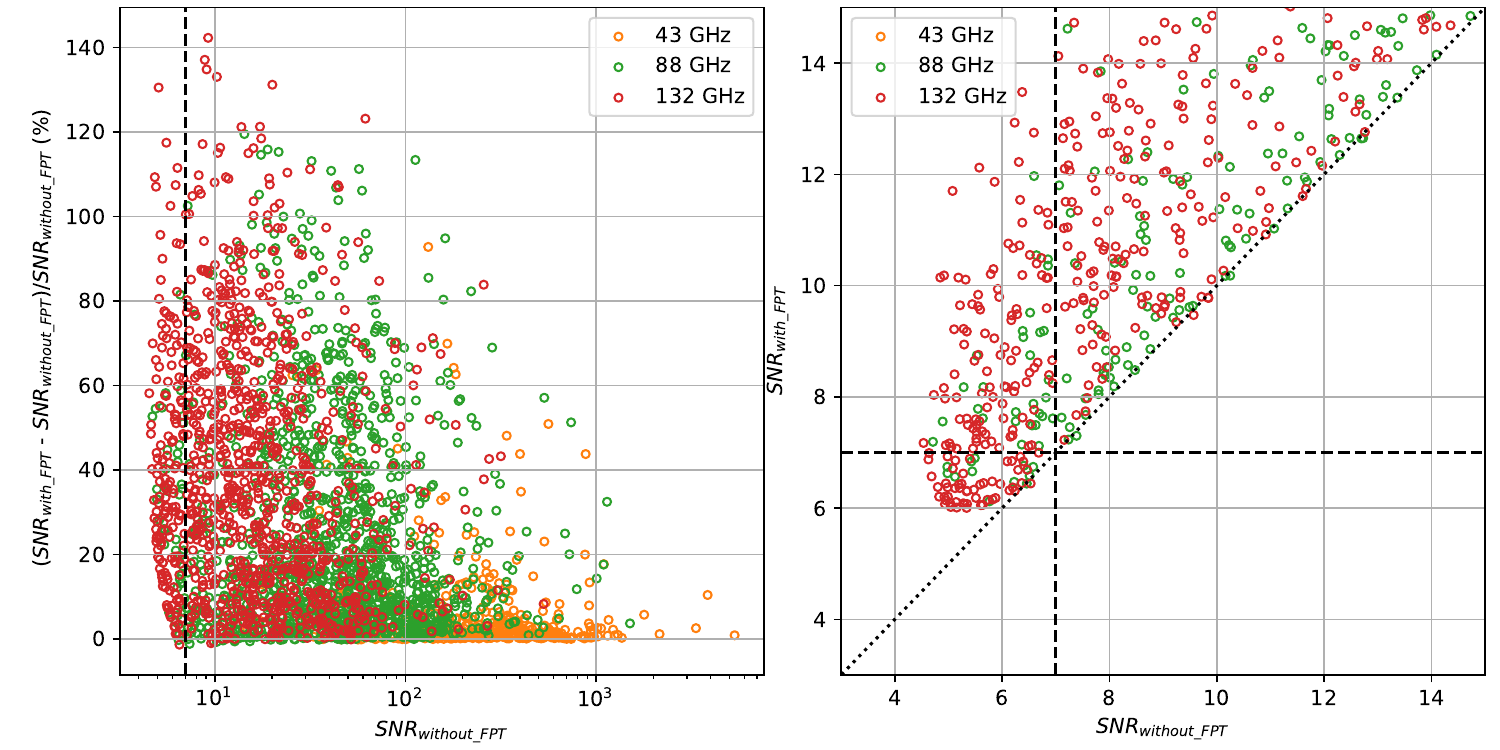}
\caption{Comparison of the SNR with and without FPT within a 2-minute scan length. The left figure shows the percentage increase in SNR after using FPT, and the right figure compares the SNR of weak sources (SNR $<$ 15). The dashed lines indicate the {\it fourfit} detection limit (SNR $=$ 7). 
For clarity, the results with an SNR below 6 using FPT are not shown.
}
\label{fig:snr}
\end{figure}

As shown in Figure \ref{fig:snr}, FPT improved the SNR of most fringes, some by more than  100$\%$ with a 2-minute scan length, resulting in a higher detection rate (over 100 observables) for weak sources at millimeter wavebands.
Note that most sources in this experiment are bright. Detection can be further improved in experiments with weaker sources and longer scan lengths.

\begin{figure}[htb!]
\epsscale{0.55}
\plotone{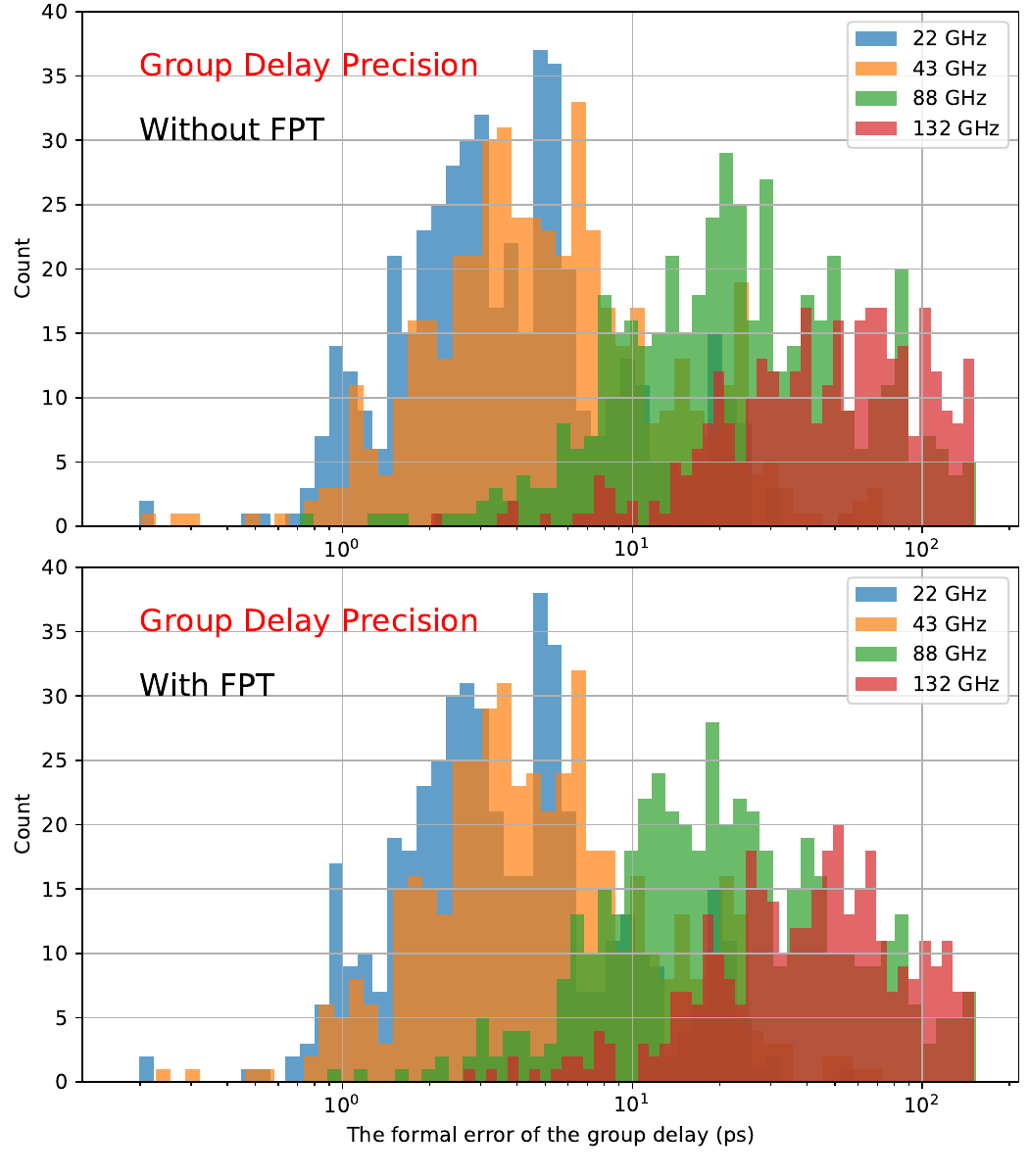}
\plotone{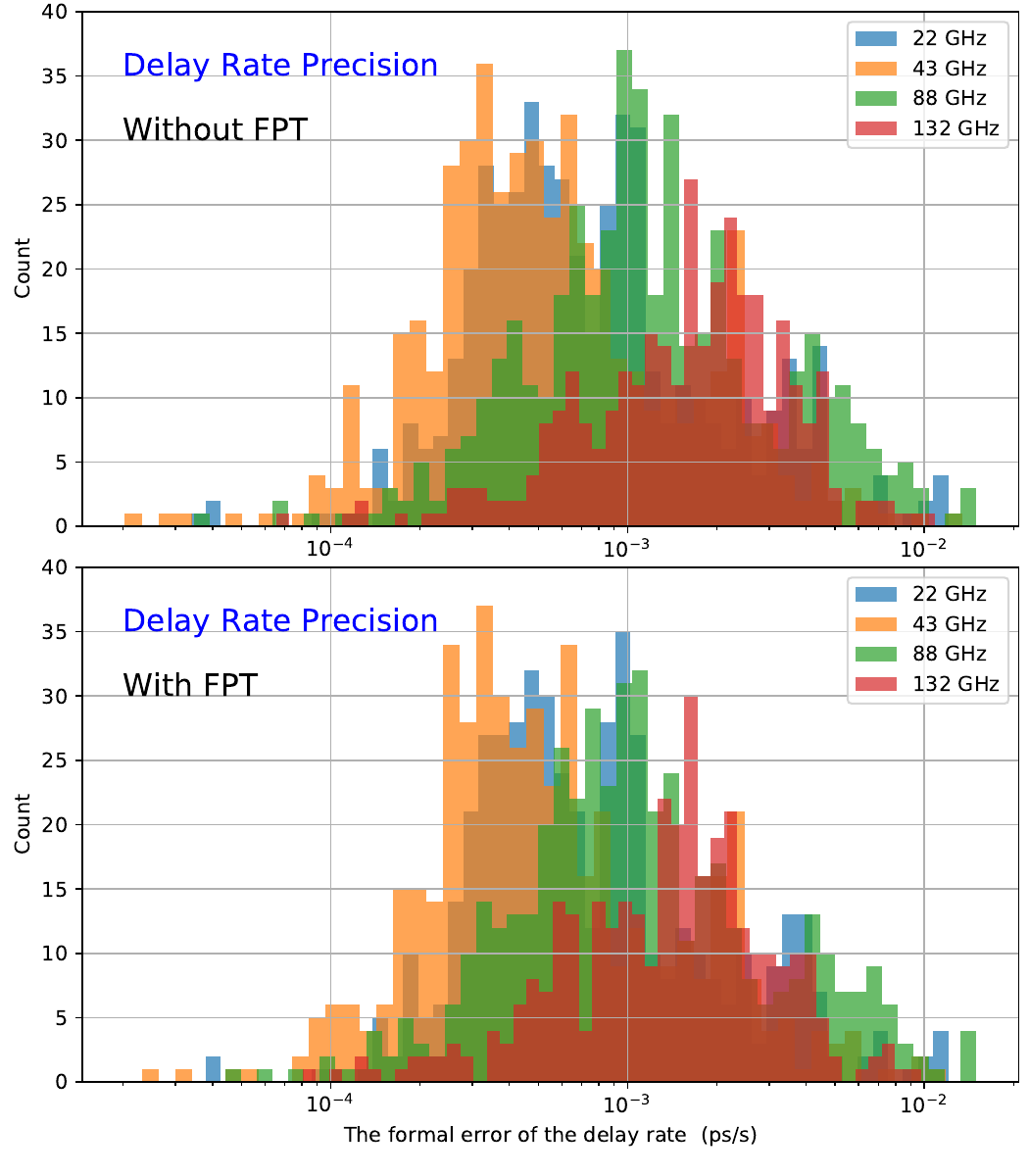}
\caption{The histogram of measurement noise for group delay and delay rate produced by {\it fourfit} on the KYS-KTN baseline. Different bands are represented by different colors.
\label{fig:err}}
\end{figure}

The theoretical uncertainties of group delays produced by {\it fourfit} are calculated using   $\sqrt{12}/(2\pi$*SNR*512 MHz) for the single 512 MHz channel data. 
Figure \ref{fig:err} shows the measurement noise of group delays on KYS-KTN baseline for each band.  The median formal errors of group delays are 3.5 ps at K band, 5.0/4.7 ps at Q band, 22.5/19.5 ps at W band, and 49.7/43.0 ps at D band  without/with FPT. Similarly, the median formal errors of delay rates are 6.6E-4 ps/s at K band, 5.0E-4/4.8E-4 ps/s at Q band, 1.1E-3/9.8E-4 ps/s at W band, and 1.7E-3/1.5E-3 ps/s at D band without/with FPT.

The closure group delay for a triangle of three stations simultaneously observing the same source is given by the sum of the three baseline group delays going around 
a closed loop of the triangle
\citep{2016AJ....152..151X}.
In this summation, the effects of station-based delays (tropospheric delays, ionospheric delays, station position errors, station thermal deformation errors, clock offset errors, cable delay errors, EOP errors, errors from pointing offsets, and so on) cancel exactly. 
The major error terms at centimeter wavebands that remained in the closure quantities, the so-called nonclosing errors, are source structure and measurement noise \citep{2018JGRB..12310162A}.  
However, in millimeter wavebands, atmospheric phase fluctuations become significant within seconds and differ at each triangle baseline, leading to errors in baseline-dependent fringe fitting when employing a 2D linear model (phase vs. time, phase vs. frequency) over a 2-minute scanning duration. As a result, uncalibrated atmospheric phase fluctuations will contribute to non-closure errors in millimeter wavebands when using {\it HOPS-fourfit}.

\begin{figure}[htb!]
\epsscale{1.0}
\plotone{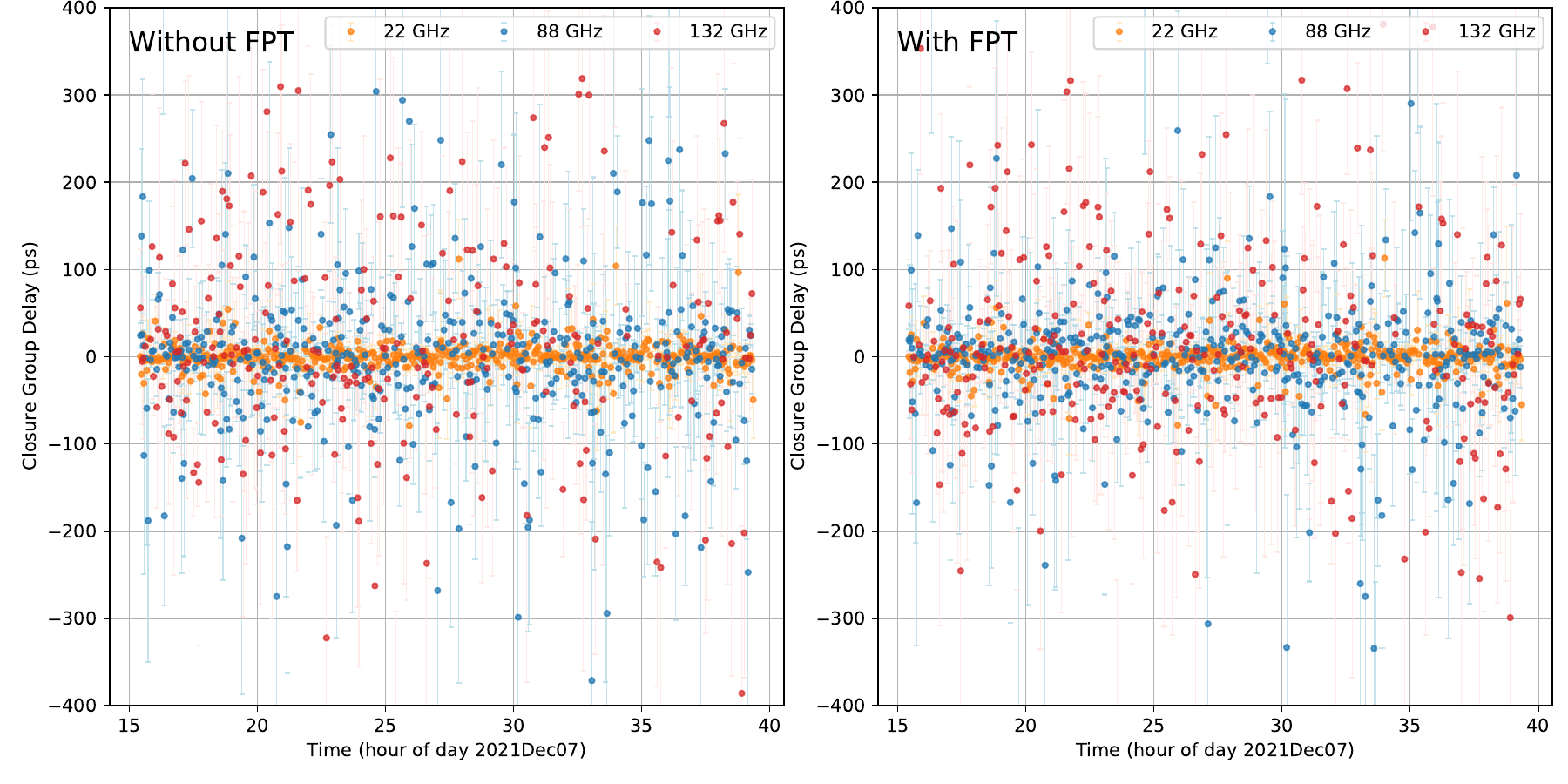}
\plotone{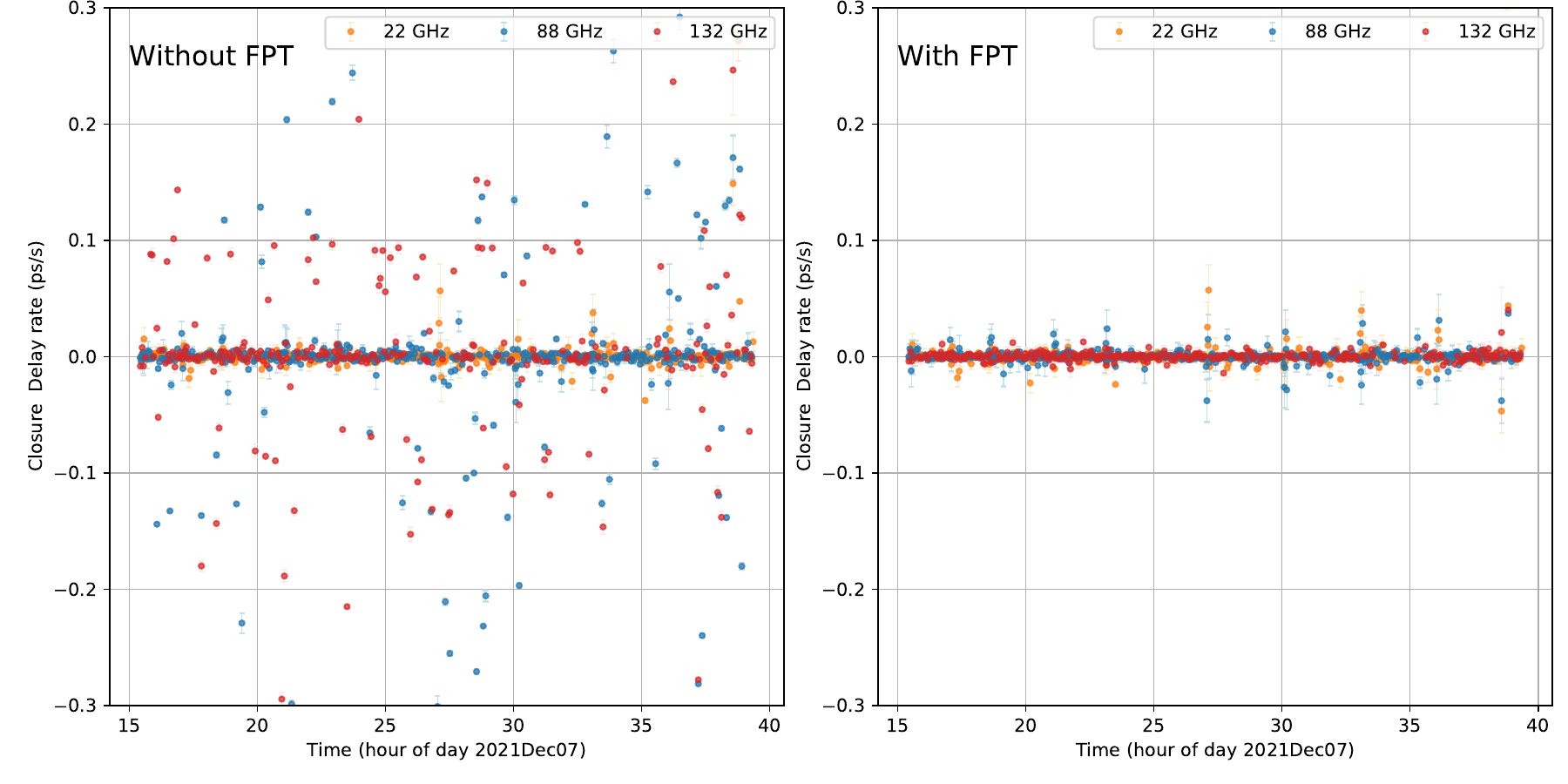}
\plotone{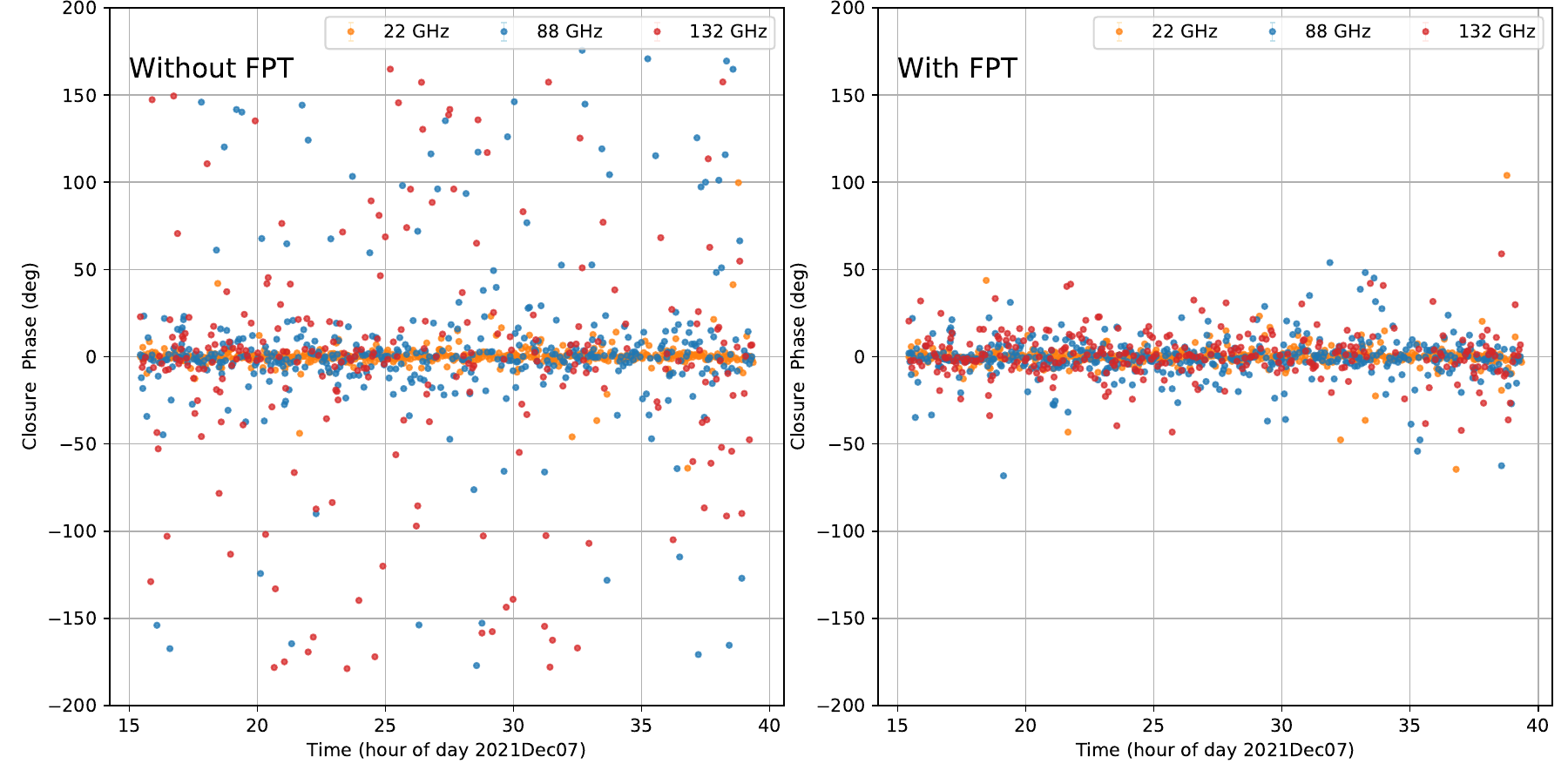}
\caption{
The closure group delay, closure delay rate, and closure phase in the KVN triangle baseline (KTN-KUS-KYS), with different frequency bands represented by dots of different colors. 
Up panels: closure group delay without/with FPT; Middle panels: closure  delay rate without/with FPT; and Bottom panels: closure phase without/with FPT. 
The 43 GHz result  is not presented  and has large uncertainties  due to the low SNR issues on KUS station.
\label{fig:clo}}
\end{figure}

Figure \ref{fig:clo} and Table \ref{tab:clo} show the closure observables in the KVN triangle baseline (KTN-KUS-KYS, with SNR $>=$ 7 on each single baseline). 
With the exception of notable errors in the Q band attributed to the KUS station issue,  the weighted standard deviations of closure group delays are 3.7 ps at K band, 19.9/15.7 ps at W band, and 47.2/36.7 ps at D band without/with FPT. These values commendably align with their corresponding formal errors, as illustrated in Figure \ref{fig:err}. Moreover, the application of FPT noticeably enhances the accuracy of the closure group delays at the W and D bands.
Similarly, the weighted standard deviations of closure delay rates are 1.7E-3 ps/s at K band, 3.5E-2/1.3E-3 ps/s at W band, and 6.1E-2/1.3E-3 ps/s at D band without/with FPT. And the weighted standard deviations of closure phases are 2.1\deg\ at K band, 24.4\deg/3.1\deg\ at W band, and 55.9\deg/4.4\deg\ at D band without/with FPT.
The FPT method significantly enhances  the closure delay rate and closure phase by over an order of magnitude at the W and D bands.

\begin{deluxetable*}{crlrrrrrrr}[htb!]
\tablecaption{Statistics of the closure observables in KVN triangle baselines \label{tab:clo}}
\tablewidth{0pt}
\tablehead{
\colhead{Band\tablenotemark{a}} & \colhead{Frequency} & \colhead{Atmospheric phase} & \colhead{Number of \tablenotemark{b}} & \colhead{Group delay} & \colhead{Group delay}  & \colhead{Delay rate} & \colhead{Delay rate}  & \colhead{Phase} & \colhead{Phase} \\
\colhead{ID} & \colhead{(MHz)} & \colhead{  calibration}  & \colhead{triangle}  & \colhead{WMean\tablenotemark{c} (ps)}& \colhead{WRMS (ps)}  & \colhead{WMean (ps/s)}& \colhead{WRMS (ps/s)} & \colhead{WMean (\deg)}& \colhead{WRMS (\deg)}
}
\startdata
K & 21984 & NO  & 481 (99.2\%)& -0.2$\pm$1.7 &  3.7 & -6.1E-5$\pm$7.9E-5 &  1.7E-3 & -0.1$\pm$0.1 & 2.1   \\
  &       &  $\phi_{Self\_K}$   & 481 (99.2\%)& -0.2$\pm$0.2 &  3.6  & -2.3E-4$\pm$1.1E-4 & 2.6E-3 & -0.2$\pm$0.1 &  2.2  \\
W & 87936 & NO  & 437 (90.1\%)& 1.4$\pm$1.0 &  19.9   & 1.1E-3$\pm$1.7E-3 &  3.5E-2 & -2.4$\pm$1.2 &  24.4  \\
  &       & $\phi_{FPT\_W}$   & 454 (93.6\%)& 1.8$\pm$0.7 &  15.7  & -3.5E-5$\pm$6.2E-5 & 1.3E-3 & -0.7$\pm$0.2 &  3.1  \\
D &131904 & NO  & 254 (52.4\%)& 0.7$\pm$3.0 &  47.2   & 1.9E-2$\pm$3.8E-3 &  6.1E-2 & 9.4$\pm$3.5 &  55.9  \\   
  &       & $\phi_{FPT\_D}$   & 301 (62.1\%)&2.1$\pm$2.1 &  36.7  & 1.4E-4$\pm$7.7E-5 & 1.3E-3 & 0.2$\pm$0.3 & 4.4   \\
\enddata
\tablenotetext{a}{The Q band result is omitted due to the presence of low SNR issues at the KUS station.
\tablenotetext{b}{Number of triangle baselines used}
\tablenotetext{c}{Weighted mean}}
\end{deluxetable*}

\subsection{The tropospheric and ionospheric effects with KVN} \label{sec:iono}

The path delay of radio waves caused by the troposphere 
is one of the major error sources. 
Modeling the tropospheric delay is generally divided into hydrostatic and wet parts, each of which is the product of the zenith delay and the corresponding mapping function dependency on elevation angle \citep{2013aesg.book.....B}. 
In {\it nuSolve}, the zenith hydrostatic delay (ZHD) is modeled as a function of the surface pressure, and the zenith wet delay  (ZWD) is calculated with the relative humidity and temperature at the surface \citep{1972GMS....15..247S,1985RaSc...20.1593D}. 
However, the uncertainty of ZWD model is far larger than ZHD model due to high spatial and temporal variability and unpredictability of the amount of water vapor.
Therefore, the residual ZWD is then parameterized as a PWL function of time (e.g., the interval of 30 min in this experiment) in the data analysis.
We compared the solutions with two different sets of priori ZHD+ZWD: one using meteorological data (pressure, relative humidity, temperature)  from the station's log files and the other using a constant value from the standard model in {\it nuSolve}. The use of meteorological data resulted in increased weighted root-mean-squares (WRMS) of the post-fitting residuals from $\sim$12 ps to $\sim$15 ps with K band data. 
In this experiment, the temperature or water vapor content at the KVN site, including their temporal changes, might not accurately reflect the conditions of the air masses above.
Therefore, we did not adopt the KVN meteorological data in the following analysis.

The KVN system  provides the
greatest spanned frequency range yet used to calibrate the ionosphere. 
The ionospheric delays of single K band or Q band geodetic observations were usually calibrated using the global vertical total electron content (TEC) map derived from global navigation satellite system (GNSS) observations provided by the International GNSS Service (IGS) with a temporal resolution of 120 minutes \citep{2010AJ....139.1695L,2020A&A...644A.159C}. 
We also tested  the regional vertical TEC map provided by the Korea Astronomy and Space Science Institute GNSS network (KASINet) \citep{https://doi.org/10.1029/2022SW003131}, which has a temporal resolution of 5 minutes and includes observing data from co-located GNSS stations at each KVN site.
The differential slant TEC (dTEC) for KVN derived with either IGS or KASINet TEC maps is only a few TEC units (TECU), except for the noon time.  One TECU corresponds to an ionospheric delay of 2.5 ps, 0.6 ps, 0.15 ps, and 0.07 ps at 22 GHz, 43 GHz, 88 GHz, and 132 GHz, respectively. Consequently, the ionospheric delays in our KVN observations are at the 10 ps level for the K band and are negligible for the Q, W, and D bands.

With the single KTN-KYS baseline, the post-fit delay residuals have a WRMS of 13.2 ps at K band and 11.6 ps at Q band. This discrepancy could potentially stem from reduced ionospheric effects at the Q band. For the W and D bands, larger WRMS values are observed due to relatively larger delay uncertainties. The utilization of either GNSS TEC maps or dual-band VLBI combinations did not yield noticeable improvements in WRMS with this data, likely due to the negligible ionospheric effects on the short baselines. 
Notably, considering that global vertical ionospheric effects typically vary from a few to dozens of TECUs \citep{Nothnagel2019}, the multi-band KVN system holds promise for obtaining ionospheric-free delays across the entire broad bandwidth from 20 to over 100 GHz with long baselines in the near future.

\subsection{Geodetic results} \label{sec:geo-results}

\begin{figure}[htb!]
\epsscale{0.55}
\plotone{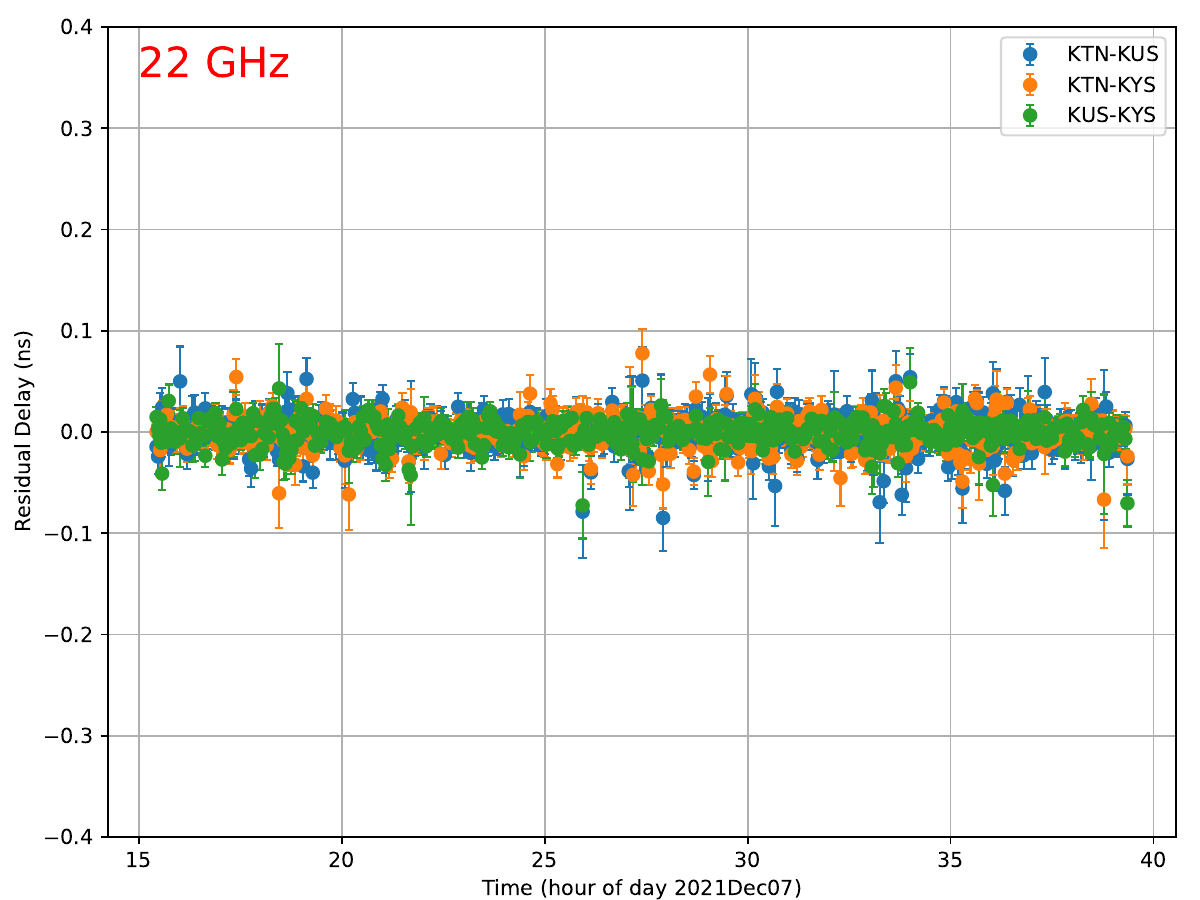}
\plotone{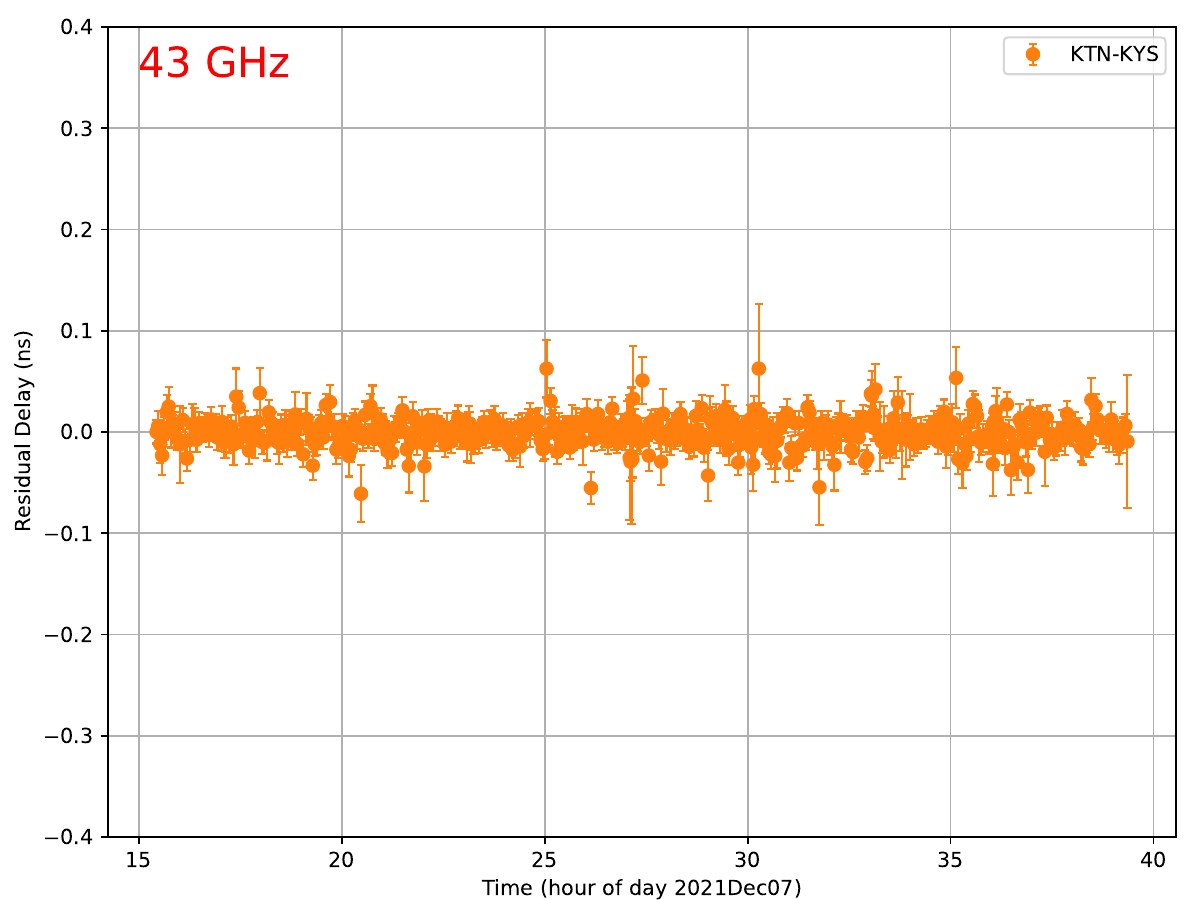}
\plotone{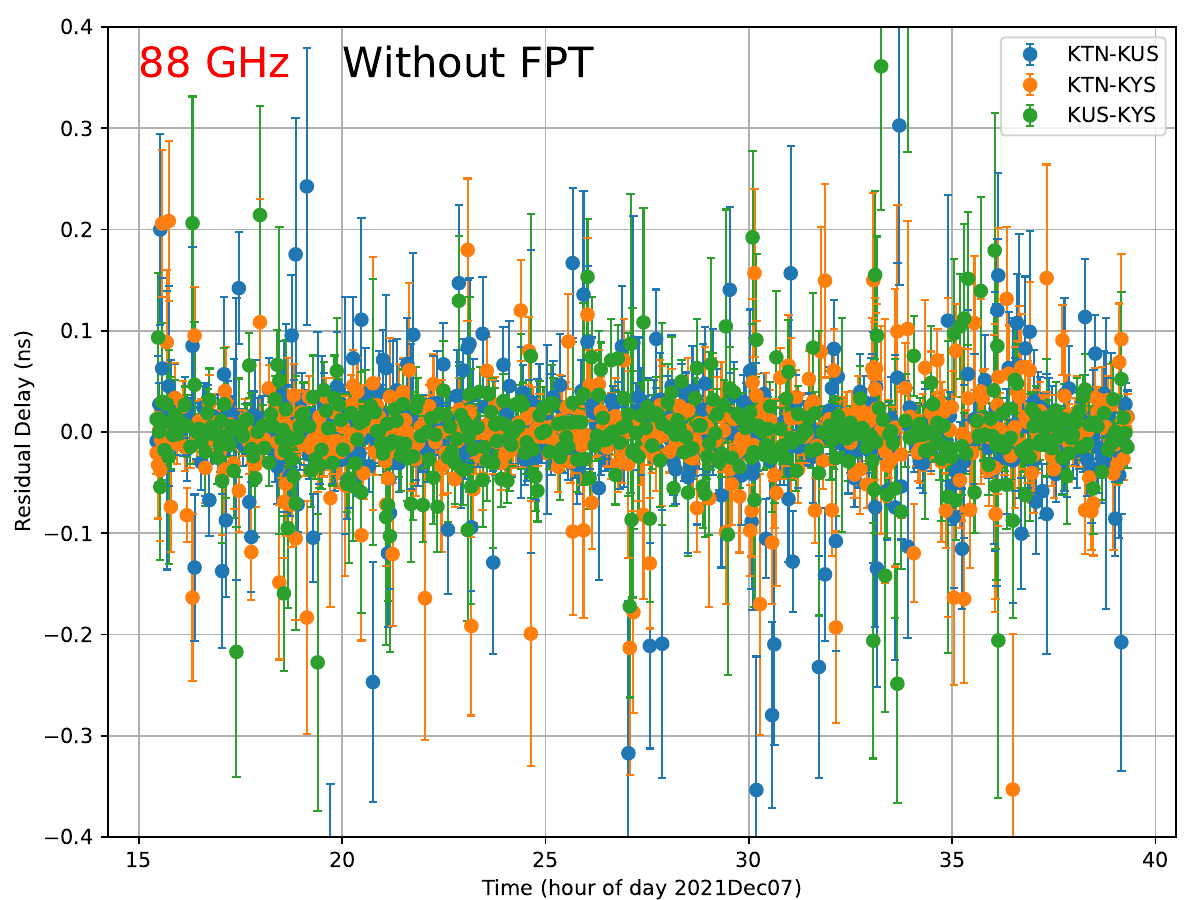}
\plotone{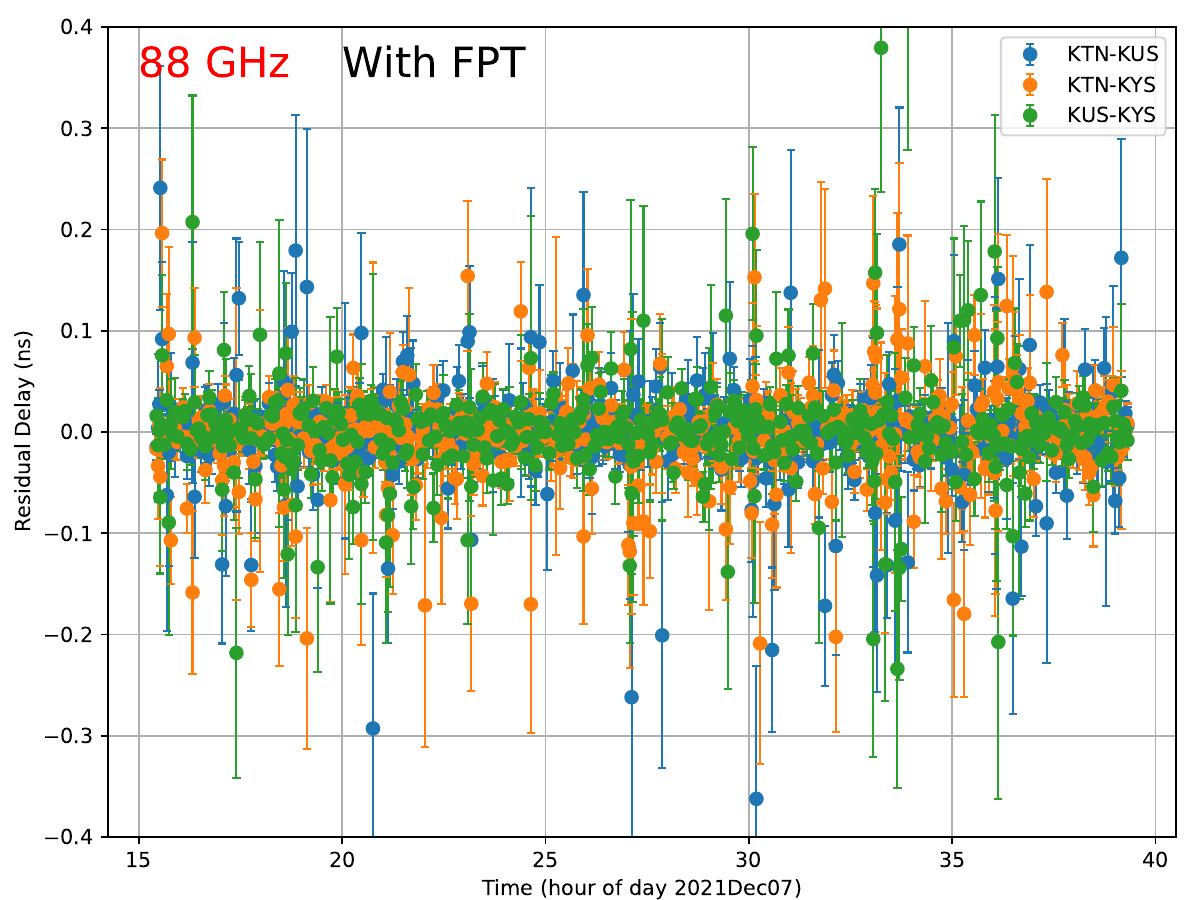}
\plotone{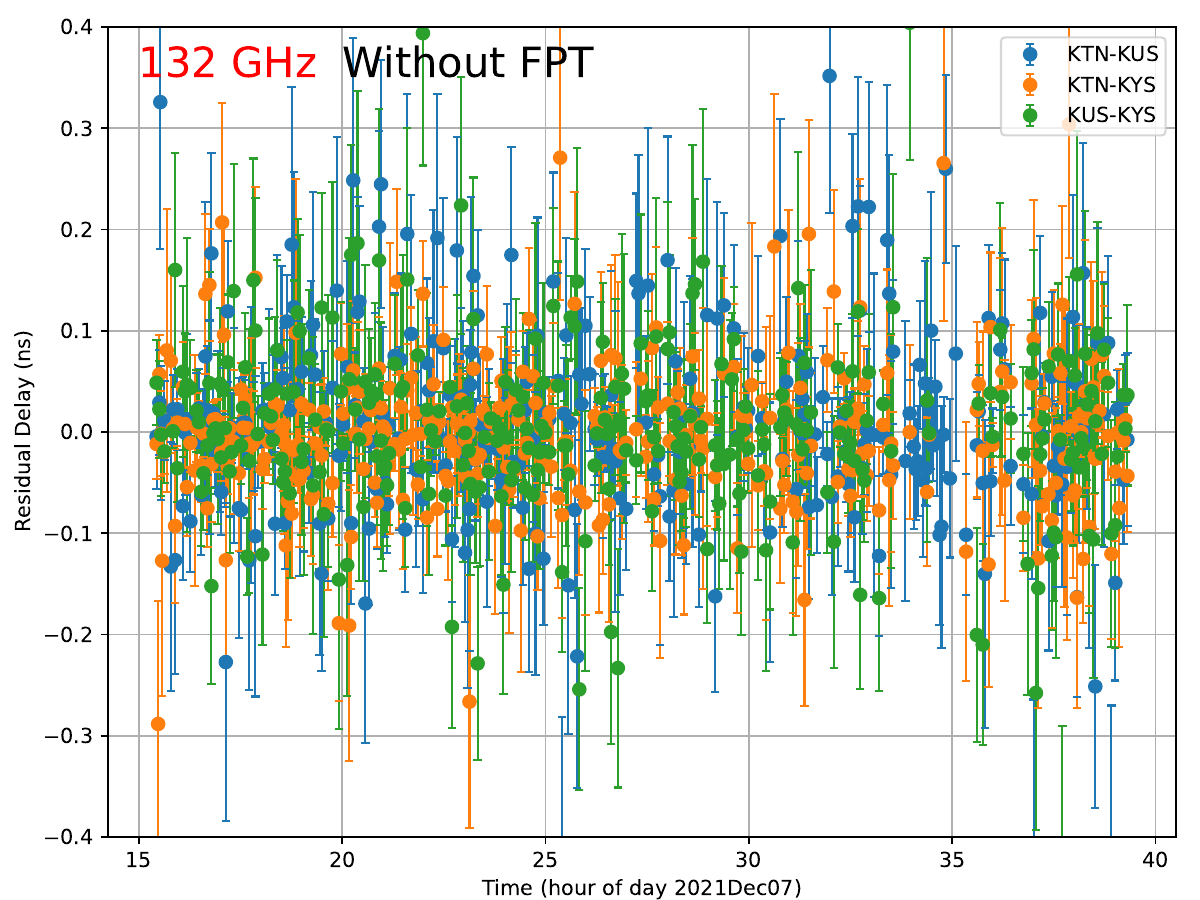}
\plotone{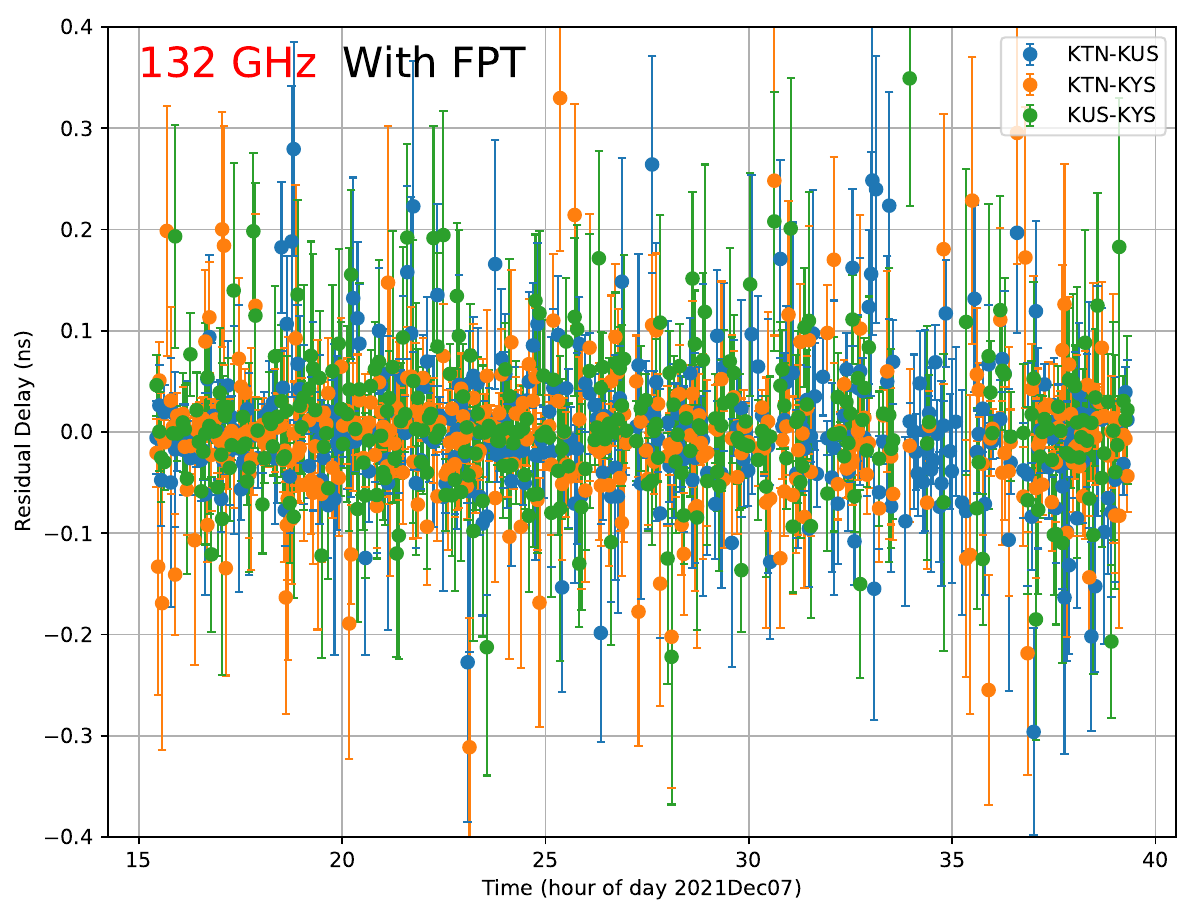}
\caption{
Post-fitting residuals at different bands }
\label{fig:rms}
\end{figure}

\begin{deluxetable*}{lclrrccccc}
\tabletypesize{\scriptsize}
\tablewidth{0pt} 
\tablecaption{Geodetic results}\label{tab:result}
\tablewidth{0pt}
\tablehead{
\colhead{Band} & \colhead{Frequency} & \colhead{Atmospheric phase} & \colhead{Recoverable} & \colhead{Used}  & \colhead{WRMS} & \colhead{chi2pdof} 
& \colhead{ KTN - KUS} & \colhead{KTN - KYS } & \colhead{KUS - KYS } \\
\colhead{ID} & \colhead{(MHz)}   & \colhead{calibration}   & \colhead{delays}    & \colhead{delays}  & \colhead{(ps)}& \colhead{}
& \colhead{ (mm) } & \colhead{ (mm)} & \colhead{ (mm)}
}
\startdata
K & 21984  &  NO                &  1447 (99.5\%)  &  1444 (99.2\%) &  12.38 & 1.00  &  358342595.57$\pm$0.78 & 476351705.73$\pm$0.82 & 304829482.93$\pm$0.71  \\        
  &        &  $\phi_{Self\_K}$     &  1447 (99.5\%)  &  1444 (99.2\%) &  12.31 & 1.00  &  358342595.63$\pm$0.78 & 476351705.73$\pm$0.81 & 304829482.85$\pm$0.71  \\   
Q\tablenotemark{a} & 42620 & NO &  484 (99.8\%)   &  483 (99.6\%)  &  11.64 & 1.00  &  -                      & 476351705.86$\pm$0.97 & - \\   
  &        &   $\phi_{FPT\_Q}$        &  484 (99.8\%)   &  483 (99.6\%)  &  11.76 & 1.00  &  -                      & 476351706.24$\pm$0.98 & - \\   
W & 87936  &  NO                &  1371 (94.2\%)  &  1369 (94.1\%) &  25.02 & 1.00  &  358342592.57$\pm$1.65 & 476351702.87$\pm$1.83 & 304829480.99$\pm$1.61  \\    
  &        &   $\phi_{FPT\_W}$        &  1397 (96.0\%)  &  1389 (95.5\%) &  20.46 & 1.00  &  358342593.85$\pm$1.26 & 476351703.33$\pm$1.53 & 304829480.73$\pm$1.37  \\     
D & 131904 &  NO                &   936 (64.3\%)  &   934 (64.2\%) &  39.34 & 1.00  &  358342599.57$\pm$3.60 & 476351707.85$\pm$3.61 & 304829482.82$\pm$4.19  \\    
  &        &   $\phi_{FPT\_D}$        &  1038 (71.3\%)  &  1032 (70.9\%) &  27.56 & 0.99  &  358342595.64$\pm$2.60 & 476351705.15$\pm$2.94 & 304829482.85$\pm$3.00  \\   
\hline  
\enddata
\tablenotetext{a}{Only with KTN-KYS baseline due to the presence of low SNR issues at the KUS station.}
\end{deluxetable*}

We performed a comparative analysis of WRMS post-fit delay residuals (pfdr) and estimated baseline lengths, as detailed in Table \ref{tab:result} and Figure \ref{fig:rms}, considering various frequency bands and the presence or absence of FPT. The baseline lengths estimated and reported in Table \ref{tab:result} are consistent with their respective uncertainties.

The post-fit delay residuals exhibit a WRMS of 12.4 ps at K band and 11.8 ps at Q band. When employing FPT, there is no noticeable difference in WRMS at K and Q bands. This suggests that the impact of atmospheric phase fluctuations at K and Q bands, particularly for baselines of less than 500 km and 2-minute scans, is minimal. This finding aligns with the similar closure group delays observed in Figure \ref{fig:clo} and Table \ref{tab:clo}.
Notably, the delay accuracy at the Q band is similar with that at the K band, and the slightly smaller WRMS at the Q band may be attributed to reduced ionospheric effects.

In the case of W and D bands, the visibility phases are significantly affected by atmospheric  turbulence. 
As depicted in Figure \ref{fig:snr}, FPT enhances fringe SNR, leading to an additional 100 delay observables at the D band. It is worth noting that the W band already boasts a very high detection rate  ($\sim 95 \%$).
Additionally, FPT reduced systematic errors in group delay and delay rate.
The WRMS of the post-fitting residuals decreased from 25.0 ps to 20.5 ps at the W band and from 39.3 ps to 27.6 ps at the D band. This indicates that the errors introduced by atmospheric phase fluctuations contributing to the WRMS are $\sim$14 ps for the W band and $\sim$28 ps for the D band in this experiment. This finding aligns with the effects observed in closure group delays in Figure \ref{fig:clo} and Table \ref{tab:clo}.
The  baseline lengths can be different at 4 mm between without and with FPT. And using FPT, the baseline lengths at W and D bands are generally closer to that of the K band. 
Therefore, calibration for atmospheric phase fluctuations is necessary in geodetic VLBI at millimeter wavebands.

The source positions estimated in geodetic analysis for this experiment achieve milliarcsecond precision, posing challenges in investigating sub-milliarcsecond frequency-dependent position offsets. Nevertheless, this can be achieved with multiple epochs, extended baselines, and/or including the SFPR method.

\section{Conclusion and Outlook}

We have successfully conducted the first simultaneous geodetic and astrometric VLBI experiment at 22/43/88/132 GHz with KVN. Our achievement includes a high detection rate of approximately $95\%$ at the W band and about  $70\%$  at the D band. Moreover, we have obtained competitive accuracy when compared to traditional centimeter wavebands, with the WRMS of the post-fitting residuals measuring 12.4 ps at K band, 11.8 ps at Q band, 20.5 ps at W band, and 27.6 ps at D band. This experiment demonstrates that the millimeter waveband can be used for geodetic applications with high precision.

For the first time, we introduced the FPT method to geodetic VLBI analysis, an approach for calibrating atmospheric phase fluctuations. 
Our results demonstrated that FPT  improves fringe detection and enhances the accuracy of delay measurements at the W and D bands. We found that atmospheric phase fluctuations, prevalent with baselines under 500 km and 2-minute scans, contribute to errors of approximately 14 ps at the W band and 28 ps at the D band. These fluctuations represent a major error source in millimeter-wave geodetic VLBI and can be effectively mitigated through the FPT method for general sources.
It is important to emphasize that FPT proves highly beneficial for geodetic mm-VLBI observations. This extends beyond merely detecting weaker sources, encompassing the precise measurement of frequency-dependent offsets in source positions at millimeter wavebands \citep{2011AJ....141..114R}.

The challenges we previously believed regarding geodetic VLBI at the millimeter waveband might be surmountable.
The high detection rate at 22/43/88/132 GHz for 82 sources has encouraged us to complete an all-sky distributed source catalog for geodetic VLBI and  mm-VLBI studies in astrophysics (e.g., \citealt{2017NewAR..79...85D}).
The number of detectable ICRF sources at K/Q/W bands is expected to exceed several hundred with KVN \citep{xu_2024_10902979}.
Typical mm-VLBI observations (above 80 GHz) require reference pointing scans to ensure accurate antenna pointing. In this experiment, such scans were not implemented due to the lack of support in {\it SKED}. Nevertheless, the results were satisfactory, owing to the high pointing accuracy of KVN. Further efforts to incorporate reference pointing scans into geodetic mm-VLBI scheduling are currently underway.
The precision of group delay with broad bandwidth synthesis from 20 to 100 GHz with KVN can be comparable to or better than VGOS, as demonstrated by a fringe in \cite{xu_2024_10902979}. 
The KVN phase-cal system is also being tested to determine the delay and phase offsets among different bands and monitor the effect of changes in ambient temperature on components of the signal chain. 
While the current KVN baseline is relatively short and limits our ability to explore the ionosphere, source structure, and frequency-dependent offsets in source positions, it's worth noting that simultaneous tri-band (K/Q/W) receivers 
are undergoing global development (e.g., \citealt{2023arXiv230604516D}).
We anticipate that we will have access to longer baselines in the coming years, enabling more comprehensive investigations in these areas as mentioned in Section \ref{sec:intro}.

\appendix 

\section{Source catalog}\label{app}

The  82 sources used in this experiment  are listed in Table \ref{tab:sou}.

\startlongtable
\tabletypesize{\scriptsize}
\begin{deluxetable*}{rrr|rrrr|rrrr}
\tablecaption{ 82 sources for geodetic VLBI at K/Q/W/D band \label{tab:sou}}
\tablewidth{0pt}
\tablehead{
\colhead{ID} & \colhead{J2000} & \colhead{IVS} & \colhead{Flux density} & \colhead{Flux density} & \colhead{Flux density} & \colhead{Flux density} & \colhead{SNR K} & \colhead{SNR Q\tablenotemark{a}} & \colhead{SNR W} & \colhead{SNR D}\\
\colhead{} & \colhead{Name} & \colhead{Name} & \colhead{K (mJy)}& \colhead{Q\tablenotemark{a} (mJy)} & \colhead{W (mJy)} & \colhead{D (mJy)} & \colhead{}& \colhead{}
}
\startdata
1  & J0006$-$0623 & 0003$-$066 & 4498$\pm$348 & 2750$\pm$99 & 947$\pm$570 & 459$\pm$162 & 827$\pm$161 & 526$\pm$134 & 73$\pm$36 & 11$\pm$11 \\ 
2  & J0019$+$7327 & 0016$+$731 & 1794$\pm$152 & 1543$\pm$62 & 834$\pm$222 & 393$\pm$140 & 357$\pm$75 & 337$\pm$45 & 62$\pm$16 & 17$\pm$9 \\ 
3  & J0102$+$5824 & 0059$+$581 & 4330$\pm$374 & 3970$\pm$270 & 2530$\pm$1099 & 795$\pm$747 & 953$\pm$127 & 993$\pm$70 & 245$\pm$45 & 90$\pm$39 \\ 
4  & J0108$+$0135 & 0106$+$013 & 4603$\pm$373 & 3740$\pm$324 & 2216$\pm$590 & 951$\pm$341 & 822$\pm$200 & 818$\pm$236 & 198$\pm$70 & 53$\pm$29 \\ 
5  & J0116$-$1136 & 0113$-$118 & 1635$\pm$133 & 1207$\pm$64 & 700$\pm$297 & 371$\pm$164 & 304$\pm$54 & 256$\pm$56 & 47$\pm$16 & 13$\pm$7 \\
6  & J0121$+$1149 & 0119$+$115 & 922$\pm$102 & 376$\pm$45 & 305$\pm$206 & 347$\pm$306 & 170$\pm$38 & 80$\pm$30 & 10$\pm$11 & 6$\pm$2 \\
7  & J0136$+$4751 & 0133$+$476 & 2589$\pm$208 & 2334$\pm$72 & 1453$\pm$312 & 748$\pm$273 & 520$\pm$79 & 525$\pm$77 & 123$\pm$30 & 41$\pm$18 \\ 
8  & J0152$+$2207 & 0149$+$218 & 526$\pm$78 & 465$\pm$3 & 311$\pm$66 & 271$\pm$79 & 105$\pm$10 & 106$\pm$19 & 24$\pm$5 & 7$\pm$2 \\
9  & J0228$+$6721 & 4C67.05 & 835$\pm$92 & 705$\pm$39 & 442$\pm$137 & 308$\pm$153 & 162$\pm$52 & 151$\pm$14 & 30$\pm$11 & 10$\pm$5 \\
10 & J0237$+$2848 & 0234$+$285 & 1752$\pm$129 & 2113$\pm$5 & 1838$\pm$343 & 1050$\pm$313 & 371$\pm$40 & 444$\pm$0 & 136$\pm$12 & 52$\pm$13 \\ 
11 & J0242$+$1101 & 0239$+$108 & 715$\pm$79 & 642$\pm$11 & 432$\pm$145 & 281$\pm$159 & 144$\pm$28 & 162$\pm$42 & 35$\pm$11 & 9$\pm$5 \\
12 & J0259$-$0019 & 0256$-$005 & 423$\pm$69 & 303$\pm$3 & 215$\pm$74 & 241$\pm$160 & 84$\pm$13 & 68$\pm$8 & 11$\pm$3 & 5$\pm$0 \\
13 & J0303$+$4716 & 0300$+$470 & 1188$\pm$100 & 905$\pm$35 & 549$\pm$149 & 344$\pm$157 & 233$\pm$31 & 176$\pm$27 & 46$\pm$8 & 16$\pm$5 \\
14 & J0325$+$2224 & 0322$+$222 & 740$\pm$78 & 689$\pm$8 & 452$\pm$147 & 351$\pm$84 & 149$\pm$20 & 156$\pm$11 & 36$\pm$5 & 11$\pm$5 \\
15 & J0336$+$3218 & NRAO140 & 1348$\pm$140 & 1563$\pm$76 & 1225$\pm$257 & 608$\pm$208 & 282$\pm$42 & 361$\pm$45 & 94$\pm$17 & 29$\pm$12 \\ 
16 & J0339$-$0146 & CTA26 & 395$\pm$103 & 432$\pm$28 & 421$\pm$94 & 386$\pm$190 & 70$\pm$22 & 84$\pm$22 & 24$\pm$7 & 7$\pm$3 \\
17 & J0422$+$0219 & 0420$+$022 & 611$\pm$83 & 458$\pm$10 & 312$\pm$149 & 314$\pm$107 & 107$\pm$24 & 90$\pm$25 & 17$\pm$6 & 5$\pm$2 \\
18 & J0422$+$5324 & 0418$+$532 & 283$\pm$77 & 232$\pm$6 & 201$\pm$93 & 245$\pm$84 & 53$\pm$17 & 44$\pm$15 & 9$\pm$3 & 5$\pm$1 \\
19 & J0423$+$4150 & 0420$+$417 & 1120$\pm$110 & 918$\pm$48 & 536$\pm$140 & 364$\pm$71 & 225$\pm$37 & 166$\pm$44 & 40$\pm$8 & 12$\pm$6 \\
20 & J0449$+$1121 & 0446$+$112 & 829$\pm$93 & 673$\pm$11 & 496$\pm$136 & 299$\pm$178 & 153$\pm$33 & 130$\pm$45 & 30$\pm$11 & 7$\pm$4 \\
21 & J0457$-$2324 & 0454$-$234 & 3350$\pm$507 & 2530$\pm$837 & 2120$\pm$278 & 1070$\pm$235 & 573$\pm$155 & 504$\pm$168 & 122$\pm$48 & 37$\pm$19 \\ 
22 & J0501$-$0159 & 0458$-$020 & 2883$\pm$237 & 2717$\pm$47 & 1955$\pm$216 & 904$\pm$227 & 564$\pm$96 & 624$\pm$121 & 153$\pm$38 & 53$\pm$23 \\ 
23 & J0530$+$1331 & 0528$+$134 & 1523$\pm$175 & 1494$\pm$22 & 1021$\pm$364 & 404$\pm$220 & 308$\pm$61 & 342$\pm$88 & 75$\pm$25 & 23$\pm$13 \\ 
24 & J0533$+$4822 & 0529$+$483 & 868$\pm$87 & 702$\pm$5 & 453$\pm$158 & 278$\pm$142 & 179$\pm$26 & 157$\pm$23 & 35$\pm$7 & 11$\pm$5 \\
25 & J0607$-$0834 & 0605$-$085 & 2204$\pm$201 & 1724$\pm$73 & 972$\pm$250 & 449$\pm$108 & 413$\pm$73 & 361$\pm$74 & 68$\pm$17 & 17$\pm$9 \\ 
26 & J0646$+$4451 & 0642$+$449 & 1817$\pm$152 & 1180$\pm$68 & 600$\pm$158 & 324$\pm$108 & 385$\pm$60 & 254$\pm$57 & 43$\pm$11 & 12$\pm$6 \\
27 & J0650$-$1637 & 0648$-$165 & 2498$\pm$249 & 2141$\pm$71 & 1517$\pm$254 & 861$\pm$174 & 440$\pm$131 & 310$\pm$137 & 76$\pm$39 & 23$\pm$17 \\ 
28 & J0725$-$0054 & 0723$-$008 & 2121$\pm$572 & 2014$\pm$57 & 1385$\pm$328 & 786$\pm$172 & 291$\pm$147 & 307$\pm$135 & 73$\pm$38 & 19$\pm$19 \\ 
29 & J0730$-$1141 & 0727$-$115 & 2619$\pm$180 & 1874$\pm$7 & 1210$\pm$200 & 656$\pm$171 & 494$\pm$62 & 383$\pm$33 & 90$\pm$8 & 31$\pm$8 \\ 
30 & J0738$+$1742 & 0735$+$178 & 867$\pm$93 & 867$\pm$3 & 788$\pm$274 & 495$\pm$132 & 167$\pm$34 & 183$\pm$46 & 56$\pm$17 & 21$\pm$11 \\ 
31 & J0808$-$0751 & 0805$-$077 & 2496$\pm$222 & 2680$\pm$59 & 2395$\pm$461 & 1556$\pm$564 & 240$\pm$39 & 253$\pm$40 & 60$\pm$10 & 19$\pm$8 \\ 
32 & J0808$+$4052 & 0805$+$410 & 1159$\pm$107 & 1019$\pm$19 & 709$\pm$239 & 392$\pm$135 & 387$\pm$80 & 381$\pm$89 & 107$\pm$35 & 30$\pm$22 \\ 
33 & J0811$+$0146 & 0808$+$019 & 696$\pm$83 & 744$\pm$7 & 680$\pm$216 & 447$\pm$103 & 127$\pm$23 & 137$\pm$26 & 39$\pm$10 & 14$\pm$7 \\
34 & J0836$-$2016 & 0834$-$201 & 398$\pm$78 & 239$\pm$6 & 250$\pm$127 & 310$\pm$162 & 68$\pm$15 & 38$\pm$9 & 8$\pm$2 & 5$\pm$0 \\
35 & J0854$+$2006 & OJ287 & 8535$\pm$694 & 8121$\pm$129 & 5995$\pm$253 & 3129$\pm$392 & 1604$\pm$303 & 1604$\pm$386 & 414$\pm$105 & 151$\pm$69 \\ 
36 & J0909$+$0121 & 0906$+$015 & 1090$\pm$103 & 961$\pm$3 & 696$\pm$121 & 475$\pm$203 & 194$\pm$31 & 174$\pm$29 & 41$\pm$8 & 13$\pm$6 \\
37 & J0921$+$6215 & 0917$+$624 & 920$\pm$100 & 830$\pm$52 & 516$\pm$130 & 355$\pm$103 & 191$\pm$29 & 188$\pm$32 & 45$\pm$11 & 17$\pm$7 \\ 
38 & J0927$+$3902 & 4C39.25 & 6298$\pm$544 & 3338$\pm$1227 & 1750$\pm$239 & 780$\pm$167 & 952$\pm$140 & 685$\pm$129 & 115$\pm$14 & 36$\pm$8 \\ 
39 & J0948$+$4039 & 0945$+$408 & 647$\pm$94 & 505$\pm$29 & 311$\pm$49 & 261$\pm$60 & 95$\pm$19 & 89$\pm$12 & 16$\pm$4 & 5$\pm$2 \\
40 & J0956$+$2515 & OK290 & 940$\pm$89 & 840$\pm$7 & 565$\pm$69 & 349$\pm$114 & 191$\pm$28 & 186$\pm$23 & 46$\pm$7 & 16$\pm$3 \\
41 & J0958$+$4725 & 0955$+$476 & 984$\pm$96 & 860$\pm$27 & 605$\pm$102 & 404$\pm$115 & 201$\pm$35 & 192$\pm$36 & 49$\pm$11 & 18$\pm$8 \\ 
42 & J1043$+$2408 & 1040$+$244 & 908$\pm$89 & 703$\pm$13 & 483$\pm$77 & 333$\pm$121 & 180$\pm$26 & 152$\pm$10 & 36$\pm$4 & 12$\pm$3 \\
43 & J1048$-$1909 & 1045$-$188 & 1068$\pm$93 & 896$\pm$15 & 646$\pm$42 & 421$\pm$59 & 198$\pm$23 & 175$\pm$17 & 45$\pm$6 & 16$\pm$4 \\
44 & J1058$+$0133 & 1055$+$018 & 5941$\pm$431 & 5160$\pm$115 & 3839$\pm$413 & 2201$\pm$507 & 1122$\pm$185 & 970$\pm$224 & 270$\pm$56 & 89$\pm$40 \\ 
45 & J1058$+$8114 & 1053$+$815 & 238$\pm$63 & 288$\pm$10 & 304$\pm$79 & 270$\pm$84 & 44$\pm$7 & 55$\pm$7 & 17$\pm$5 & 5$\pm$3 \\
46 & J1146$+$3958 & 1144$+$402 & 1355$\pm$105 & 964$\pm$115 & 598$\pm$106 & 356$\pm$85 & 285$\pm$42 & 247$\pm$31 & 50$\pm$9 & 19$\pm$6 \\ 
47 & J1147$-$0724 & 1145$-$071 & 642$\pm$92 & 375$\pm$73 & 218$\pm$100 & 278$\pm$170 & 117$\pm$21 & 68$\pm$24 & 8$\pm$3 & 5$\pm$0 \\
48 & J1159$+$2914 & 1156$+$295 & 5792$\pm$451 & 5931$\pm$213 & 5906$\pm$805 & 2916$\pm$359 & 1156$\pm$174 & 1276$\pm$156 & 365$\pm$71 & 149$\pm$50 \\ 
49 & J1215$-$1731 & 1213$-$172 & 2384$\pm$277 & 2095$\pm$72 & 1372$\pm$264 & 697$\pm$191 & 432$\pm$120 & 386$\pm$128 & 86$\pm$37 & 22$\pm$16 \\ 
50 & J1230$+$1223 & 3C274 & 1644$\pm$169 & 1281$\pm$123 & 1055$\pm$250 & 673$\pm$177 & 318$\pm$46 & 300$\pm$50 & 85$\pm$15 & 32$\pm$14 \\ 
51 & J1256$-$0547 & 3C279 & 24891$\pm$1725 & 22193$\pm$824 & 15208$\pm$1651 & 7325$\pm$1443 & 4492$\pm$715 & 4254$\pm$800 & 1120$\pm$268 & 356$\pm$118 \\ 
52 & J1310$+$3220 & 1308$+$326 & 2444$\pm$194 & 2420$\pm$86 & 1725$\pm$234 & 889$\pm$220 & 519$\pm$78 & 612$\pm$74 & 149$\pm$19 & 46$\pm$13 \\ 
53 & J1337$-$1257 & 1334$-$127 & 2662$\pm$207 & 2056$\pm$594 & 1967$\pm$250 & 1046$\pm$236 & 472$\pm$121 & 449$\pm$156 & 114$\pm$56 & 36$\pm$28 \\ 
54 & J1357$+$1919 & 1354$+$195 & 1741$\pm$141 & 1648$\pm$128 & 1018$\pm$248 & 480$\pm$177 & 357$\pm$48 & 422$\pm$50 & 87$\pm$17 & 25$\pm$13 \\ 
55 & J1419$+$5423 & 1418$+$546 & 622$\pm$98 & 538$\pm$54 & 355$\pm$80 & 271$\pm$152 & 130$\pm$25 & 114$\pm$24 & 27$\pm$8 & 8$\pm$4 \\
56 & J1459$+$7140 & 3C309.1 & 335$\pm$107 & 279$\pm$32 & 197$\pm$53 & 269$\pm$104 & 62$\pm$22 & 50$\pm$6 & 7$\pm$2 & 5$\pm$0 \\
57 & J1504$+$1029 & 1502$+$106 & 831$\pm$86 & 721$\pm$21 & 450$\pm$102 & 305$\pm$108 & 182$\pm$24 & 182$\pm$10 & 41$\pm$6 & 16$\pm$4 \\
58 & J1512$-$0905 & 1510$-$089 & 3643$\pm$245 & 2669$\pm$340 & 1656$\pm$260 & 771$\pm$195 & 694$\pm$115 & 593$\pm$113 & 126$\pm$32 & 40$\pm$15 \\ 
59 & J1540$+$1447 & 1538$+$149 & 974$\pm$97 & 901$\pm$4 & 753$\pm$63 & 546$\pm$141 & 163$\pm$19 & 153$\pm$0 & 42$\pm$5 & 13$\pm$5 \\
60 & J1549$+$0237 & 1546$+$027 & 2464$\pm$162 & 1568$\pm$59 & 693$\pm$295 & 275$\pm$177 & 478$\pm$99 & 395$\pm$91 & 47$\pm$24 & 8$\pm$7 \\
61 & J1613$+$3412 & 1611$+$343 & 2125$\pm$191 & 1228$\pm$63 & 539$\pm$271 & 362$\pm$110 & 436$\pm$53 & 240$\pm$13 & 37$\pm$14 & 13$\pm$4 \\ 
62 & J1625$-$2527 & 1622$-$253 & 1014$\pm$103 & 756$\pm$18 & 489$\pm$58 & 439$\pm$141 & 169$\pm$24 & 126$\pm$18 & 26$\pm$4 & 6$\pm$1 \\
63 & J1638$+$5720 & 1637$+$574 & 926$\pm$85 & 987$\pm$66 & 765$\pm$187 & 463$\pm$142 & 202$\pm$24 & 219$\pm$12 & 65$\pm$12 & 26$\pm$9 \\ 
64 & J1642$-$0621 & 1639$-$062 & 1374$\pm$123 & 1104$\pm$42 & 696$\pm$112 & 371$\pm$152 & 268$\pm$48 & 240$\pm$49 & 52$\pm$11 & 15$\pm$6 \\ 
65 & J1658$+$0741 & 1655$+$077 & 1342$\pm$107 & 1167$\pm$41 & 741$\pm$157 & 385$\pm$85 & 280$\pm$41 & 290$\pm$31 & 64$\pm$10 & 22$\pm$7 \\ 
66 & J1733$-$1304 & NRAO530 & 3360$\pm$123 & 1789$\pm$59 & 1076$\pm$136 & 493$\pm$89 & 612$\pm$95 & 334$\pm$73 & 76$\pm$16 & 17$\pm$5 \\ 
67 & J1734$+$3857 & 1732$+$389 & 1221$\pm$106 & 1229$\pm$61 & 843$\pm$278 & 480$\pm$213 & 262$\pm$35 & 291$\pm$30 & 70$\pm$8 & 25$\pm$8 \\ 
68 & J1751$+$0939 & 1749$+$096 & 2768$\pm$199 & 2505$\pm$81 & 1633$\pm$208 & 826$\pm$193 & 616$\pm$80 & 653$\pm$46 & 152$\pm$20 & 51$\pm$12 \\ 
69 & J1753$+$2848 & 1751$+$288 & 2235$\pm$173 & 1736$\pm$60 & 1002$\pm$230 & 451$\pm$206 & 430$\pm$65 & 378$\pm$57 & 76$\pm$12 & 27$\pm$6 \\ 
70 & J1800$+$7828 & 1803$+$784 & 2099$\pm$179 & 1739$\pm$113 & 1073$\pm$212 & 579$\pm$152 & 405$\pm$63 & 337$\pm$44 & 83$\pm$17 & 27$\pm$12 \\ 
71 & J1824$+$5651 & 1823$+$568 & 926$\pm$164 & 787$\pm$63 & 513$\pm$129 & 355$\pm$96 & 199$\pm$39 & 179$\pm$25 & 41$\pm$13 & 14$\pm$7 \\ 
72 & J1849$+$6705 & 1849$+$670 & 1604$\pm$282 & 1257$\pm$267 & 616$\pm$319 & 377$\pm$153 & 306$\pm$121 & 268$\pm$91 & 56$\pm$22 & 16$\pm$9 \\ 
73 & J1911$-$2006 & 1908$-$201 & 2374$\pm$231 & 1769$\pm$206 & 1136$\pm$163 & 563$\pm$132 & 393$\pm$107 & 282$\pm$114 & 68$\pm$31 & 18$\pm$9 \\ 
74 & J2000$-$1748 & 1958$-$179 & 1803$\pm$163 & 1783$\pm$97 & 1232$\pm$203 & 612$\pm$179 & 309$\pm$54 & 332$\pm$57 & 86$\pm$25 & 22$\pm$12 \\ 
75 & J2115$+$2933 & 2113$+$293 & 849$\pm$75 & 626$\pm$8 & 369$\pm$173 & 253$\pm$109 & 168$\pm$26 & 139$\pm$23 & 27$\pm$5 & 6$\pm$3 \\
76 & J2134$-$0153 & 2131$-$021 & 1484$\pm$166 & 1088$\pm$149 & 642$\pm$227 & 386$\pm$164 & 285$\pm$63 & 219$\pm$79 & 45$\pm$18 & 9$\pm$9 \\
77 & J2139$+$1423 & 2136$+$141 & 1434$\pm$118 & 811$\pm$42 & 347$\pm$99 & 249$\pm$137 & 234$\pm$61 & 152$\pm$43 & 22$\pm$7 & 5$\pm$1 \\
78 & J2152$+$1734 & 2150$+$173 & 300$\pm$70 & 249$\pm$8 & 246$\pm$103 & 238$\pm$288 & 41$\pm$9 & 41$\pm$10 & 9$\pm$3 & 5$\pm$0 \\
79 & J2229$-$0832 & 2227$-$088 & 2460$\pm$186 & 2570$\pm$132 & 1542$\pm$577 & 537$\pm$337 & 471$\pm$76 & 547$\pm$96 & 119$\pm$33 & 32$\pm$20 \\ 
80 & J2232$+$1143 & CTA102 & 3792$\pm$285 & 3733$\pm$1679 & 3140$\pm$619 & 755$\pm$498 & 749$\pm$135 & 1061$\pm$256 & 283$\pm$97 & 73$\pm$61 \\ 
81 & J2236$+$2828 & 2234$+$282 & 1940$\pm$144 & 1870$\pm$100 & 1152$\pm$762 & 664$\pm$159 & 388$\pm$64 & 413$\pm$77 & 94$\pm$25 & 27$\pm$16 \\ 
82 & J2327$+$0940 & 2325$+$093 & 1328$\pm$123 & 1371$\pm$95 & 792$\pm$438 & 505$\pm$121 & 262$\pm$51 & 312$\pm$72 & 59$\pm$20 & 16$\pm$11 \\ 
\enddata
\tablecomments{The flux density values are obtained from KVN baseline  ($<500$ km) amplitude using {\it AIPS}, while the  SNR with 2-min scans is produced by {\it fourfit}.  These values (weighted averages and standard deviations) are provided as a reference for conducting  observations and may not be accurate for AGN astrophysics studies. 
The different  standard deviations observed in the baseline amplitudes and SNRs may be attributed to variations in weather conditions, elevation angles, and pointing accuracy.
}
\tablenotetext{a}{Q band is only from the KYS-KTN baseline. }
\end{deluxetable*}

\begin{acknowledgments}
This work utilized the KVN under the EAVN program. We are grateful to all staff members in KVN and EAVN who helped to operate the array.
The KVN and a high-performance computing cluster are facilities operated by the KASI (Korea Astronomy and Space Science Institute). The KVN observations and correlations are supported through the high-speed network connections among the KVN sites provided by the KREONET (Korea Research Environment Open NETwork), which is managed and operated by the KISTI (Korea Institute of Science and Technology Information).
The presented figures were generated using Matplotlib \citep{Hunter:2007} and Astropy \citep{2013A&A...558A..33A}.
This research was supported by the National Research Council of Science \& Technology(NST) grant by the Korea government (MSIT) (No. CAP22061-000).
BZ was supported by the National Natural Science Foundation of China (Grant No. U2031212 and U1831136),
and Shanghai Astronomical Observatory, Chinese Academy of Sciences (Grant No. N2020-06-19-005). SX thanks Dr. John Barrett and Dr. Daniel Hoak for solving the technical problems with {\it HOPS}, 
and thanks to Dr. Sergei Bolotin for the guide on {\it nuSolve}.  
\end{acknowledgments}
 
\vspace{5mm} 
\facilities{KVN, EAVN} 
\software{
{\it HOPS} \citep{2022Galax..10..119H},
{\it nuSolve} \citep{2014ivs..conf..253B},
{\it SKED} \citep{gipson2018sked},
{\it SCHED} \citep{walker2022sched},
{\it DiFX}  \citep{2011PASP..123..275D},
{\it AIPS} \citep{2003ASSL..285..109G},
{\it ParselTongue} \citep{2006ASPC..351..497K}.
} 

\bibliography{ref}{} 

\begin{thebibliography}{}
\expandafter\ifx\csname natexlab\endcsname\relax\def\natexlab#1{#1}\fi
\providecommand{\url}[1]{\href{#1}{#1}}
\providecommand{\dodoi}[1]{doi:~\href{http://doi.org/#1}{\nolinkurl{#1}}}
\providecommand{\doeprint}[1]{\href{http://ascl.net/#1}{\nolinkurl{http://ascl.net/#1}}}
\providecommand{\doarXiv}[1]{\href{https://arxiv.org/abs/#1}{\nolinkurl{https://arxiv.org/abs/#1}}}

\bibitem[{{Akiyama} {et~al.}(2022){Akiyama}, {Algaba}, {An}, {Asada}, {Asanok}, {Byun}, {Chanapote}, {Chen}, {Chen}, {Cheng}, {Chibueze}, {Cho}, {Cho}, {Chung}, {Cui}, {Cui}, {Doi}, {Dong}, {Fujisawa}, {Gou}, {Guo}, {Hada}, {Hagiwara}, {Hirota}, {Hodgson}, {Honma}, {Imai}, {Jaroenjittichai}, {Jiang}, {Jiang}, {Jiang}, {Jike}, {Jung}, {Jung}, {Kawaguchi}, {Kim}, {Kim}, {Kim}, {Kim}, {Kim}, {Kim}, {Kino}, {Kobayashi}, {Koyama}, {Kramer}, {Lee}, {Lee}, {Lee}, {Lee}, {Li}, {Li}, {Li}, {Li}, {Liu}, {Liu}, {Lu}, {Motogi}, {Nakamura}, {Niinuma}, {Oh}, {Oh}, {Oh}, {Oh}, {Oyama}, {Park}, {Poshyachinda}, {Ro}, {Roh}, {Rujopakarn}, {Sakai}, {Sawada-Satoh}, {Shen}, {Shibata}, {Sohn}, {Soonthornthum}, {Sugiyama}, {Sun}, {Takamura}, {Tanabe}, {Tazaki}, {Trippe}, {Wajima}, {Wang}, {Wang}, {Wang}, {Wang}, {Xia}, {Xu}, {Yan}, {Yang}, {Yeom}, {Yi}, {Yi}, {Yonekura}, {Yoon}, {Yu}, {Yuan}, {Yun}, {Zhang}, {Zhang}, {Zhang}, {Zhao}, {Zhao}, \& {Zhong}}]{2022Galax..10..113A}
{Akiyama}, K., {Algaba}, J.-C., {An}, T., {et~al.} 2022, Galaxies, 10, 113, \dodoi{10.3390/galaxies10060113}

\bibitem[{{Anderson} \& {Xu}(2018)}]{2018JGRB..12310162A}
{Anderson}, J.~M., \& {Xu}, M.~H. 2018, Journal of Geophysical Research (Solid Earth), 123, 10,162, \dodoi{10.1029/2018JB015550}

\bibitem[{{Astropy Collaboration} {et~al.}(2013){Astropy Collaboration}, {Robitaille}, {Tollerud}, {Greenfield}, {Droettboom}, {Bray}, {Aldcroft}, {Davis}, {Ginsburg}, {Price-Whelan}, {Kerzendorf}, {Conley}, {Crighton}, {Barbary}, {Muna}, {Ferguson}, {Grollier}, {Parikh}, {Nair}, {Unther}, {Deil}, {Woillez}, {Conseil}, {Kramer}, {Turner}, {Singer}, {Fox}, {Weaver}, {Zabalza}, {Edwards}, {Azalee Bostroem}, {Burke}, {Casey}, {Crawford}, {Dencheva}, {Ely}, {Jenness}, {Labrie}, {Lim}, {Pierfederici}, {Pontzen}, {Ptak}, {Refsdal}, {Servillat}, \& {Streicher}}]{2013A&A...558A..33A}
{Astropy Collaboration}, {Robitaille}, T.~P., {Tollerud}, E.~J., {et~al.} 2013, \aap, 558, A33, \dodoi{10.1051/0004-6361/201322068}

\bibitem[{{Blackburn} {et~al.}(2019){Blackburn}, {Chan}, {Crew}, {Fish}, {Issaoun}, {Johnson}, {Wielgus}, {Akiyama}, {Barrett}, {Bouman}, {Cappallo}, {Chael}, {Janssen}, {Lonsdale}, \& {Doeleman}}]{2019ApJ...882...23B}
{Blackburn}, L., {Chan}, C.-k., {Crew}, G.~B., {et~al.} 2019, \apj, 882, 23, \dodoi{10.3847/1538-4357/ab328d}

\bibitem[{{Blandford} \& {K{\"o}nigl}(1979)}]{1979ApJ...232...34B}
{Blandford}, R.~D., \& {K{\"o}nigl}, A. 1979, \apj, 232, 34, \dodoi{10.1086/157262}

\bibitem[{{Bohm} \& {Schuh}(2013)}]{2013aesg.book.....B}
{Bohm}, J., \& {Schuh}, H. 2013, {Atmospheric Effects in Space Geodesy}, \dodoi{10.1007/978-3-642-36932-2}

\bibitem[{{Bolotin} {et~al.}(2023){Bolotin}, {Baver}, {B{\'e}rub{\'e}}, \& {Gipson}}]{2023ivs..conf..159B}
{Bolotin}, S., {Baver}, K., {B{\'e}rub{\'e}}, M., \& {Gipson}, J. 2023, in International VLBI Service for Geodesy and Astrometry 2022 General Meeting Proceedings, ed. K.~L. {Armstrong}, D.~{Behrend}, \& K.~D. {Baver}, 159--163

\bibitem[{{Bolotin} {et~al.}(2014){Bolotin}, {Baver}, {Gipson}, {Gordon}, \& {MacMillan}}]{2014ivs..conf..253B}
{Bolotin}, S., {Baver}, K., {Gipson}, J., {Gordon}, D., \& {MacMillan}, D. 2014, in International VLBI Service for Geodesy and Astrometry 2014 General Meeting Proceedings: ''VGOS: The New VLBI Network, 253--257

\bibitem[{{Bolotin} {et~al.}(2016){Bolotin}, {Baver}, {Gipson}, {Gordon}, \& {MacMillan}}]{2016ivs..conf..222B}
{Bolotin}, S., {Baver}, K., {Gipson}, J., {Gordon}, D., \& {MacMillan}, D. 2016, in New Horizons with VGOS, ed. D.~{Behrend}, K.~D. {Baver}, \& K.~L. {Armstrong}, 222--224

\bibitem[{{Charlot}(2022)}]{Charlot2022}
{Charlot}, P. 2022, in IAU General Assembly.
\newblock \url{https://www.iau.org/static/science/scientific_bodies/divisions/a/2022/Div-A-3_9-1_Patrick_Charlot.pdf}

\bibitem[{{Charlot} {et~al.}(2020){Charlot}, {Jacobs}, {Gordon}, {Lambert}, {de Witt}, {B{\"o}hm}, {Fey}, {Heinkelmann}, {Skurikhina}, {Titov}, {Arias}, {Bolotin}, {Bourda}, {Ma}, {Malkin}, {Nothnagel}, {Mayer}, {MacMillan}, {Nilsson}, \& {Gaume}}]{2020A&A...644A.159C}
{Charlot}, P., {Jacobs}, C.~S., {Gordon}, D., {et~al.} 2020, \aap, 644, A159, \dodoi{10.1051/0004-6361/202038368}

\bibitem[{{Davis} {et~al.}(1985){Davis}, {Herring}, {Shapiro}, {Rogers}, \& {Elgered}}]{1985RaSc...20.1593D}
{Davis}, J.~L., {Herring}, T.~A., {Shapiro}, I.~I., {Rogers}, A.~E.~E., \& {Elgered}, G. 1985, Radio Science, 20, 1593, \dodoi{10.1029/RS020i006p01593}

\bibitem[{{de Witt} {et~al.}(2023){de Witt}, {Jacobs}, {Gordon}, {Bietenholz}, {Nickola}, \& {Bertarini}}]{2023AJ....165..139D}
{de Witt}, A., {Jacobs}, C.~S., {Gordon}, D., {et~al.} 2023, \aj, 165, 139, \dodoi{10.3847/1538-3881/aca012}

\bibitem[{{Deller} {et~al.}(2011){Deller}, {Brisken}, {Phillips}, {Morgan}, {Alef}, {Cappallo}, {Middelberg}, {Romney}, {Rottmann}, {Tingay}, \& {Wayth}}]{2011PASP..123..275D}
{Deller}, A.~T., {Brisken}, W.~F., {Phillips}, C.~J., {et~al.} 2011, \pasp, 123, 275, \dodoi{10.1086/658907}

\bibitem[{{Dodson} {et~al.}(2017){Dodson}, {Rioja}, {Jung}, {Gom{\'e}z}, {Bujarrabal}, {Moscadelli}, {Miller-Jones}, {Tetarenko}, \& {Sivakoff}}]{2017NewAR..79...85D}
{Dodson}, R., {Rioja}, M.~J., {Jung}, T., {et~al.} 2017, \nar, 79, 85, \dodoi{10.1016/j.newar.2017.09.003}

\bibitem[{{Dodson} {et~al.}(2023){Dodson}, {Garc{\'\i}a-Mir{\'o}}, {Giroletti}, {Jung}, {Lindqvist}, {Lobanov}, {Rioja}, {Ros}, {Savolainen}, {Sohn}, {Zensus}, \& {Zhao}}]{2023arXiv230604516D}
{Dodson}, R., {Garc{\'\i}a-Mir{\'o}}, C., {Giroletti}, M., {et~al.} 2023, arXiv e-prints, arXiv:2306.04516, \dodoi{10.48550/arXiv.2306.04516}

\bibitem[{{Fish}(2015)}]{Fish2015}
{Fish}, V. 2015, in Mm-VLBI Data Processing Workshop.
\newblock \url{https://www.jive.eu/mm-vlbi2015/Documents/Fish_tutorial.pdf}

\bibitem[{{Fomalont} {et~al.}(2009){Fomalont}, {Kopeikin}, {Lanyi}, \& {Benson}}]{2009ApJ...699.1395F}
{Fomalont}, E., {Kopeikin}, S., {Lanyi}, G., \& {Benson}, J. 2009, \apj, 699, 1395, \dodoi{10.1088/0004-637X/699/2/1395}

\bibitem[{{Gaia Collaboration} {et~al.}(2022){Gaia Collaboration}, {Klioner}, {Lindegren}, {Mignard}, {Hern{\'a}ndez}, {Ramos-Lerate}, {Bastian}, {Biermann}, {Bombrun}, {de Torres}, {Gerlach}, {Geyer}, {Hilger}, {Hobbs}, {Lammers}, {McMillan}, {Steidelm{\"u}ller}, {Teyssier}, {Raiteri}, {Bartolom{\'e}}, {Bernet}, {Casta{\~n}eda}, {Clotet}, {Davidson}, {Fabricius}, {Garralda Torres}, {Gonz{\'a}lez-Vidal}, {Portell}, {Rowell}, {Torra}, {Torra}, {Brown}, {Vallenari}, {Prusti}, {de Bruijne}, {Arenou}, {Babusiaux}, {Creevey}, {Ducourant}, {Evans}, {Eyer}, {Guerra}, {Hutton}, {Jordi}, {Luri}, {Panem}, {Pourbaix}, {Randich}, {Sartoretti}, {Soubiran}, {Tanga}, {Walton}, {Bailer-Jones}, {Drimmel}, {Jansen}, {Katz}, {Lattanzi}, {van Leeuwen}, {Bakker}, {Cacciari}, {De Angeli}, {Fouesneau}, {Fr{\'e}mat}, {Galluccio}, {Guerrier}, {Heiter}, {Masana}, {Messineo}, {Mowlavi}, {Nicolas}, {Nienartowicz}, {Pailler}, {Panuzzo}, {Riclet}, {Roux}, {Seabroke}, {Sordo}, {Th{\'e}venin}, {Gracia-Abril}, {Altmann}, {Andrae}, {Audard},
  {Bellas-Velidis}, {Benson}, {Berthier}, {Blomme}, {Burgess}, {Busonero}, {Busso}, {C{\'a}novas}, {Carry}, {Cellino}, {Cheek}, {Clementini}, {Damerdji}, {de Teodoro}, {Nu{\~n}ez Campos}, {Delchambre}, {Dell'Oro}, {Esquej}, {Fern{\'a}ndez-Hern{\'a}ndez}, {Fraile}, {Garabato}, {Garc{\'\i}a-Lario}, {Gosset}, {Haigron}, {Halbwachs}, {Hambly}, {Harrison}, {Hestroffer}, {Hodgkin}, {Holl}, {Jan{\ss}en}, {Jevardat de Fombelle}, {Jordan}, {Krone-Martins}, {Lanzafame}, {L{\"o}ffler}, {Marchal}, {Marrese}, {Moitinho}, {Muinonen}, {Osborne}, {Pancino}, {Pauwels}, {Recio-Blanco}, {Reyl{\'e}}, {Riello}, {Rimoldini}, {Roegiers}, {Rybizki}, {Sarro}, {Siopis}, {Smith}, {Sozzetti}, {Utrilla}, {van Leeuwen}, {Abbas}, {{\'A}brah{\'a}m}, {Abreu Aramburu}, {Aerts}, {Aguado}, {Ajaj}, {Aldea-Montero}, {Altavilla}, {{\'A}lvarez}, {Alves}, {Anderson}, {Anglada Varela}, {Antoja}, {Baines}, {Baker}, {Balaguer-N{\'u}{\~n}ez}, {Balbinot}, {Balog}, {Barache}, {Barbato}, {Barros}, {Barstow}, {Bassilana}, {Bauchet}, {Becciani},
  {Bellazzini}, {Berihuete}, {Bertone}, {Bianchi}, {Binnenfeld}, {Blanco-Cuaresma}, {Boch}, {Bossini}, {Bouquillon}, {Bragaglia}, {Bramante}, {Breedt}, {Bressan}, {Brouillet}, {Brugaletta}, {Bucciarelli}, {Burlacu}, {Butkevich}, {Buzzi}, {Caffau}, {Cancelliere}, {Cantat-Gaudin}, {Carballo}, {Carlucci}, {Carnerero}, {Carrasco}, {Casamiquela}, {Castellani}, {Castro-Ginard}, {Chaoul}, {Charlot}, {Chemin}, {Chiaramida}, {Chiavassa}, {Chornay}, {Comoretto}, {Contursi}, {Cooper}, {Cornez}, {Cowell}, {Crifo}, {Cropper}, {Crosta}, {Crowley}, {Dafonte}, {Dapergolas}, {David}, {de Laverny}, {De Luise}, {De March}, {De Ridder}, {de Souza}, {del Peloso}, {del Pozo}, {Delbo}, {Delgado}, {Delisle}, {Demouchy}, {Dharmawardena}, {Diakite}, {Diener}, {Distefano}, {Dolding}, {Enke}, {Fabre}, {Fabrizio}, {Faigler}, {Fedorets}, {Fernique}, {Fienga}, {Figueras}, {Fournier}, {Fouron}, {Fragkoudi}, {Gai}, {Garcia-Gutierrez}, {Garcia-Reinaldos}, {Garc{\'\i}a-Torres}, {Garofalo}, {Gavel}, {Gavras}, {Giacobbe}, {Gilmore}, {Girona},
  {Giuffrida}, {Gomel}, {Gomez}, {Gonz{\'a}lez-N{\'u}{\~n}ez}, {Gonz{\'a}lez-Santamar{\'\i}a}, {Granvik}, {Guillout}, {Guiraud}, {Guti{\'e}rrez-S{\'a}nchez}, {Guy}, {Hatzidimitriou}, {Hauser}, {Haywood}, {Helmer}, {Helmi}, {Sarmiento}, {Hidalgo}, {H{\l}adczuk}, {Holland}, {Huckle}, {Jardine}, {Jasniewicz}, {Jean-Antoine Piccolo}, {Jim{\'e}nez-Arranz}, {Juaristi Campillo}, {Julbe}, {Karbevska}, {Kervella}, {Khanna}, {Kordopatis}, {Korn}, {K{\'o}sp{\'a}l}, {Kostrzewa-Rutkowska}, {Kruszy{\'n}ska}, {Kun}, {Laizeau}, {Lambert}, {Lanza}, {Lasne}, {Le Campion}, {Lebreton}, {Lebzelter}, {Leccia}, {Leclerc}, {Lecoeur-Taibi}, {Liao}, {Licata}, {Lindstr{\o}m}, {Lister}, {Livanou}, {Lobel}, {Lorca}, {Loup}, {Madrero Pardo}, {Magdaleno Romeo}, {Managau}, {Mann}, {Manteiga}, {Marchant}, {Marconi}, {Marcos}, {Santos}, {Mar{\'\i}n Pina}, {Marinoni}, {Marocco}, {Marshall}, {Polo}, {Mart{\'\i}n-Fleitas}, {Marton}, {Mary}, {Masip}, {Massari}, {Mastrobuono-Battisti}, {Mazeh}, {Messina}, {Michalik}, {Millar}, {Mints}, {Molina},
  {Molinaro}, {Moln{\'a}r}, {Monari}, {Mongui{\'o}}, {Montegriffo}, {Montero}, {Mor}, {Mora}, {Morbidelli}, {Morel}, {Morris}, {Muraveva}, {Murphy}, {Musella}, {Nagy}, {Noval}, {Oca{\~n}a}, {Ogden}, {Ordenovic}, {Osinde}, {Pagani}, {Pagano}, {Palaversa}, {Palicio}, {Pallas-Quintela}, {Panahi}, {Payne-Wardenaar}, {Pe{\~n}alosa Esteller}, {Penttil{\"a}}, {Pichon}, {Piersimoni}, {Pineau}, {Plachy}, {Plum}, {Poggio}, {Pr{\v{s}}a}, {Pulone}, {Racero}, {Ragaini}, {Rainer}, {Rambaux}, {Ramos}, {Re Fiorentin}, {Regibo}, {Richards}, {Diaz}, {Ripepi}, {Riva}, {Rix}, {Rixon}, {Robichon}, {Robin}, {Robin}, {Roelens}, {Rogues}, {Rohrbasser}, {Romero-G{\'o}mez}, {Royer}, {Ruz Mieres}, {Rybicki}, {Sadowski}, {S{\'a}ez N{\'u}{\~n}ez}, {Sagrist{\`a} Sell{\'e}s}, {Sahlmann}, {Salguero}, {Samaras}, {Sanchez Gimenez}, {Sanna}, {Santove{\~n}a}, {Sarasso}, {Schultheis}, {Sciacca}, {Segol}, {Segovia}, {S{\'e}gransan}, {Semeux}, {Shahaf}, {Siddiqui}, {Siebert}, {Siltala}, {Silvelo}, {Slezak}, {Slezak}, {Smart}, {Snaith}, {Solano},
  {Solitro}, {Souami}, {Souchay}, {Spagna}, {Spina}, {Spoto}, {Steele}, {Stephenson}, {S{\"u}veges}, {Surdej}, {Szabados}, {Szegedi-Elek}, {Taris}, {Taylor}, {Teixeira}, {Tolomei}, {Tonello}, {Torralba Elipe}, {Trabucchi}, {Tsounis}, {Turon}, {Ulla}, {Unger}, {Vaillant}, {van Dillen}, {van Reeven}, {Vanel}, {Vecchiato}, {Viala}, {Vicente}, {Voutsinas}, {Weiler}, {Wevers}, {Wyrzykowski}, {Yoldas}, {Yvard}, {Zhao}, {Zorec}, {Zucker}, \& {Zwitter}}]{2022A&A...667A.148G}
{Gaia Collaboration}, {Klioner}, S.~A., {Lindegren}, L., {et~al.} 2022, \aap, 667, A148, \dodoi{10.1051/0004-6361/202243483}

\bibitem[{Gipson(2018)}]{gipson2018sked}
Gipson, J. 2018, SKED VLBI Scheduling Software. Goddard Space FLight Center

\bibitem[{{Greisen}(2003)}]{2003ASSL..285..109G}
{Greisen}, E.~W. 2003, in Astrophysics and Space Science Library, Vol. 285, Information Handling in Astronomy - Historical Vistas, ed. A.~{Heck}, 109, \dodoi{10.1007/0-306-48080-8\_7}

\bibitem[{{Hada} {et~al.}(2011){Hada}, {Doi}, {Kino}, {Nagai}, {Hagiwara}, \& {Kawaguchi}}]{2011Natur.477..185H}
{Hada}, K., {Doi}, A., {Kino}, M., {et~al.} 2011, \nat, 477, 185, \dodoi{10.1038/nature10387}

\bibitem[{{Han} {et~al.}(2013){Han}, {Lee}, {Kang}, {Oh}, {Byun}, {Je}, {Chung}, {Wi}, {Song}, {Kang}, {Lee}, {Kim}, {Sasao}, {Goldsmith}, \& {Wylde}}]{2013PASP..125..539H}
{Han}, S.-T., {Lee}, J.-W., {Kang}, J., {et~al.} 2013, \pasp, 125, 539, \dodoi{10.1086/671125}

\bibitem[{{Hoak} {et~al.}(2022){Hoak}, {Barrett}, {Crew}, \& {Pfeiffer}}]{2022Galax..10..119H}
{Hoak}, D., {Barrett}, J., {Crew}, G., \& {Pfeiffer}, V. 2022, Galaxies, 10, 119, \dodoi{10.3390/galaxies10060119}

\bibitem[{Hunter(2007)}]{Hunter:2007}
Hunter, J.~D. 2007, Computing in Science \& Engineering, 9, 90, \dodoi{10.1109/MCSE.2007.55}

\bibitem[{Jeong {et~al.}(2022)Jeong, Lee, Jang, Kil, Kim, Kwak, Kim, Hong, \& Choi}]{https://doi.org/10.1029/2022SW003131}
Jeong, S.-H., Lee, W.~K., Jang, S., {et~al.} 2022, Space Weather, 20, e2022SW003131, \dodoi{https://doi.org/10.1029/2022SW003131}

\bibitem[{{Jung}(2018)}]{2018evn..confE.104J}
{Jung}, T. 2018, in 14th European VLBI Network Symposium \& Users Meeting (EVN 2018), 104

\bibitem[{{Jung} {et~al.}(2011){Jung}, {Sohn}, {Kobayashi}, {Sasao}, {Hirota}, {Kameya}, {Choi}, \& {Chung}}]{2011PASJ...63..375J}
{Jung}, T., {Sohn}, B.~W., {Kobayashi}, H., {et~al.} 2011, \pasj, 63, 375, \dodoi{10.1093/pasj/63.2.375}

\bibitem[{{Jung} {et~al.}(2015){Jung}, {Dodson}, {Han}, {Rioja}, {Byun}, {Honma}, {Stevens}, {de Vincente}, \& {Sohn}}]{2015JKAS...48..277J}
{Jung}, T., {Dodson}, R., {Han}, S.-T., {et~al.} 2015, Journal of Korean Astronomical Society, 48, 277, \dodoi{10.5303/JKAS.2015.48.5.277}

\bibitem[{{Kettenis} {et~al.}(2006){Kettenis}, {van Langevelde}, {Reynolds}, \& {Cotton}}]{2006ASPC..351..497K}
{Kettenis}, M., {van Langevelde}, H.~J., {Reynolds}, C., \& {Cotton}, B. 2006, in Astronomical Society of the Pacific Conference Series, Vol. 351, Astronomical Data Analysis Software and Systems XV, ed. C.~{Gabriel}, C.~{Arviset}, D.~{Ponz}, \& S.~{Enrique}, 497

\bibitem[{{Koryukova} {et~al.}(2022){Koryukova}, {Pushkarev}, {Plavin}, \& {Kovalev}}]{2022MNRAS.515.1736K}
{Koryukova}, T.~A., {Pushkarev}, A.~B., {Plavin}, A.~V., \& {Kovalev}, Y.~Y. 2022, \mnras, 515, 1736, \dodoi{10.1093/mnras/stac189810.48550/arXiv.2201.04359}

\bibitem[{{Lanyi} {et~al.}(2010){Lanyi}, {Boboltz}, {Charlot}, {Fey}, {Fomalont}, {Geldzahler}, {Gordon}, {Jacobs}, {Ma}, {Naudet}, {Romney}, {Sovers}, \& {Zhang}}]{2010AJ....139.1695L}
{Lanyi}, G.~E., {Boboltz}, D.~A., {Charlot}, P., {et~al.} 2010, \aj, 139, 1695, \dodoi{10.1088/0004-6256/139/5/1695}

\bibitem[{{Lee} {et~al.}(2017){Lee}, {Sohn}, {Jung}, {Byun}, \& {Lee}}]{2017ApJS..228...22L}
{Lee}, J.~A., {Sohn}, B.~W., {Jung}, T., {Byun}, D.-Y., \& {Lee}, J.~W. 2017, \apjs, 228, 22, \dodoi{10.3847/1538-4365/228/2/22}

\bibitem[{{Lee} {et~al.}(2014){Lee}, {Petrov}, {Byun}, {Kim}, {Jung}, {Song}, {Oh}, {Roh}, {Je}, {Wi}, {Sohn}, {Oh}, {Kim}, {Yeom}, {Chung}, {Kang}, {Han}, {Lee}, {Kim}, {Chung}, {Kim}, {Ryoung Kim}, {Kang}, \& {Cho}}]{2014AJ....147...77L}
{Lee}, S.-S., {Petrov}, L., {Byun}, D.-Y., {et~al.} 2014, \aj, 147, 77, \dodoi{10.1088/0004-6256/147/4/77}

\bibitem[{{MacMillan}(1995)}]{1995GeoRL..22.1041M}
{MacMillan}, D.~S. 1995, \grl, 22, 1041, \dodoi{10.1029/95GL00887}

\bibitem[{{Niell} {et~al.}(2018){Niell}, {Barrett}, {Burns}, {Cappallo}, {Corey}, {Derome}, {Eckert}, {Elosegui}, {McWhirter}, {Poirier}, {Rajagopalan}, {Rogers}, {Ruszczyk}, {SooHoo}, {Titus}, {Whitney}, {Behrend}, {Bolotin}, {Gipson}, {Gordon}, {Himwich}, \& {Petrachenko}}]{2018RaSc...53.1269N}
{Niell}, A., {Barrett}, J., {Burns}, A., {et~al.} 2018, Radio Science, 53, 1269, \dodoi{10.1029/2018RS006617}

\bibitem[{{Niell} {et~al.}(2021){Niell}, {Barrett}, {Cappallo}, {Corey}, {Elosegui}, {Mondal}, {Rajagopalan}, {Ruszczyk}, \& {Titus}}]{2021JGeod..95...65N}
{Niell}, A.~E., {Barrett}, J.~P., {Cappallo}, R.~J., {et~al.} 2021, Journal of Geodesy, 95, 65, \dodoi{10.1007/s00190-021-01505-9}

\bibitem[{Nothnagel(2019)}]{Nothnagel2019}
Nothnagel, A. 2019, Very Long Baseline Interferometry, ed. W.~Freeden \& R.~Rummel (Berlin, Heidelberg: Springer Berlin Heidelberg), 1--58, \dodoi{10.1007/978-3-662-46900-2_110-1}

\bibitem[{{Petrov}(2024)}]{2024arXiv240408800P}
{Petrov}, L. 2024, arXiv e-prints, arXiv:2404.08800, \dodoi{10.48550/arXiv.2404.08800}

\bibitem[{{Rioja} \& {Dodson}(2011)}]{2011AJ....141..114R}
{Rioja}, M., \& {Dodson}, R. 2011, \aj, 141, 114, \dodoi{10.1088/0004-6256/141/4/114}

\bibitem[{{Rioja} \& {Dodson}(2020)}]{2020A&ARv..28....6R}
{Rioja}, M.~J., \& {Dodson}, R. 2020, \aapr, 28, 6, \dodoi{10.1007/s00159-020-00126-z}

\bibitem[{{Rioja} {et~al.}(2015){Rioja}, {Dodson}, {Jung}, \& {Sohn}}]{2015AJ....150..202R}
{Rioja}, M.~J., {Dodson}, R., {Jung}, T., \& {Sohn}, B.~W. 2015, \aj, 150, 202, \dodoi{10.1088/0004-6256/150/6/202}

\bibitem[{{Saastamoinen}(1972)}]{1972GMS....15..247S}
{Saastamoinen}, J. 1972, in The Use of Artificial Satellites for Geodesy, Vol.~15, 247, \dodoi{10.1029/GM015p0247}

\bibitem[{{Sovers} {et~al.}(1998){Sovers}, {Fanselow}, \& {Jacobs}}]{1998RvMP...70.1393S}
{Sovers}, O.~J., {Fanselow}, J.~L., \& {Jacobs}, C.~S. 1998, Reviews of Modern Physics, 70, 1393, \dodoi{10.1103/RevModPhys.70.1393}

\bibitem[{Walker(2022)}]{walker2022sched}
Walker, R. 2022, The sched user manual

\bibitem[{{Whitney} {et~al.}(2022){Whitney}, {Cappallo}, {Aldrich}, {Anderson}, {Bos}, {Casse}, {Goodman}, {Parsley}, {Pogrebenko}, {Schilizzi}, \& {Smythe}}]{2022ascl.soft05019W}
{Whitney}, A.~R., {Cappallo}, R., {Aldrich}, W., {et~al.} 2022, {HOPS: Haystack Observatory Postprocessing System}, Astrophysics Source Code Library, record ascl:2205.019.
\newblock \doeprint{2205.019}

\bibitem[{{Xu} {et~al.}(2021{\natexlab{a}}){Xu}, {Anderson}, {Heinkelmann}, {Lunz}, {Schuh}, \& {Wang}}]{2021JGeod..95...51X}
{Xu}, M.~H., {Anderson}, J.~M., {Heinkelmann}, R., {et~al.} 2021{\natexlab{a}}, Journal of Geodesy, 95, 51, \dodoi{10.1007/s00190-021-01496-7}

\bibitem[{{Xu} {et~al.}(2016){Xu}, {Heinkelmann}, {Anderson}, {Mora-Diaz}, {Schuh}, \& {Wang}}]{2016AJ....152..151X}
{Xu}, M.~H., {Heinkelmann}, R., {Anderson}, J.~M., {et~al.} 2016, \aj, 152, 151, \dodoi{10.3847/0004-6256/152/5/151}

\bibitem[{{Xu} {et~al.}(2022){Xu}, {Savolainen}, {Anderson}, {Kareinen}, {Zubko}, {Lunz}, \& {Schuh}}]{2022A&A...663A..83X}
{Xu}, M.~H., {Savolainen}, T., {Anderson}, J.~M., {et~al.} 2022, \aap, 663, A83, \dodoi{10.1051/0004-6361/202140840}

\bibitem[{Xu {et~al.}(2024)Xu, Jung, \& Byun}]{xu_2024_10902979}
Xu, S., Jung, T., \& Byun, D.-Y. 2024, {Geodetic and Astrometric VLBI at K/Q/W/D Bands with the KVN},  Zenodo, \dodoi{10.5281/zenodo.10902979}

\bibitem[{{Xu} {et~al.}(2021{\natexlab{b}}){Xu}, {Jike}, {Jung}, {Shu}, {Cui}, {Melnikov}, {McCallum}, {Yi}, {Zhang}, {Sakai}, {He}, {Imai}, {Kawaguchi}, {Sakai}, {Oh}, {Jiang}, {Xu}, \& {Wang}}]{2021evga.conf...71X}
{Xu}, S., {Jike}, T., {Jung}, T., {et~al.} 2021{\natexlab{b}}, in 25th European VLBI Group for Geodesy and Astrometry Working Meeting, ed. R.~{Haas}, Vol.~25, 71--73

\end{thebibliography}

\end{CJK*}
\end{document}